\documentclass[journal]{IEEEtran}

\usepackage{nomencl}
\usepackage{amssymb}
\makenomenclature

\usepackage{xcolor,soul,framed} 
\usepackage{cite}
\usepackage{textcomp}
\colorlet{shadecolor}{yellow}
\usepackage[pdftex]{graphicx}
\graphicspath{{../pdf/}{../jpeg/}}

\usepackage{etoolbox}
\renewcommand\nomgroup[1]{%
  \item[\bfseries
  \ifstrequal{#1}{A}{Indices and Sets}{%
  \ifstrequal{#1}{B}{Parameters}{%
  \ifstrequal{#1}{C}{Variables}{}}}%
]}
\DeclareGraphicsExtensions{.pdf,.jpeg,.png}

\usepackage[cmex10]{amsmath}
\usepackage{array}
\usepackage{mdwmath}
\usepackage{mdwtab}
\usepackage{eqparbox}
\usepackage{url}

\newcounter{MYtempeqncnt}

\begin{document}
\bstctlcite{IEEEexample:BSTcontrol}
    \title{Incorporating flexibility and resilience demand into capacity market considering the guidance on generation investment}
  \author{Yunpeng~Xiao,~\IEEEmembership{Senior Member,~IEEE,}
      Hui~Guo, Wenqi~Wu,~\IEEEmembership{Graduate Student Member,~IEEE,} \\
      Xiuli Wang,~\IEEEmembership{Senior Member,~IEEE,}
      and~Xifan~Wang,~\IEEEmembership{Life Fellow,~IEEE}   \thanks{This work is supported by the National Key Research and Development Program of China “Key Technology of Integrated Regulation and System Optimization of Coal Power and New Energy” under Grant No. 2022YFB2403200. \textit{(Corresponding author: Yunpeng Xiao.)}}
   \thanks{The authors are with the School of Electrical Engineering, Xi’an Jiaotong University, Xi’an 710049, China (e-mail: ypxiao@xjtu.edu.cn).}}
\maketitle

\begin{abstract}
The capacity market provides economic guidance for generation investment and ensures the adequacy of generation capability for power systems. With the rapidly increasing proportion of renewable energy, the adequacy of flexibility and resilience becomes more crucial for the secure operation of power systems. In this context, this paper incorporates the flexibility and resilience demand into the capacity market by formulating the capacity demand curves for ramping capability, inertia and recovery capabilities besides the generation capability. The guidance on generation investment of the capacity market is also taken into account by solving the generation investment equilibrium among generation companies (Gencos) with a Nash-Cournot model employing an equivalent quadratic programming formulation. The overall problem is established as a tri-level game and an iterative algorithm is devised to formulate the capacity demand curves in the upper level based on Genco’s investment acquired from the middle and lower levels. The case study further demonstrates that to incorporate flexibility and resilience demand into the capacity market could stimulate proper generation investment and ensure the adequacy of flexibility and resilience in power systems.
\end{abstract}


%
\IEEEpeerreviewmaketitle

\nomenclature[A]{\(d\in D\)}{Index and set of the segments of the demand curve}
\nomenclature[A]{\(e\in E\)}{Index and set of ESs}
\nomenclature[A]{\(i,i^{\prime}\in I\)}{Index and set of areas $i$ and $i^{\prime}$}
\nomenclature[A]{\(j\in J\)}{Index and set of Genco $j$}
\nomenclature[A]{\(k\in K\)}{Index and set of units}
\nomenclature[A]{\(l\in \Omega _{j}^{L}\)}{Index and set of transmission lines}
\nomenclature[A]{\(s\in {{\Omega }_{s}}\)}{Index and set of photovoltaic units}
\nomenclature[A]{\(t\in T\)}{Index and set of periods}
\nomenclature[A]{\(w\in {{\Omega }_{w}}\)}{Index and set of wind power units}
\nomenclature[A]{\(\omega\)}{Index of scenarios}
\nomenclature[A]{\(\Omega_i^j\)}{association of Genco $j$ and areas $i$}
\nomenclature[B]{\(C_k\)}{Cost for power generation}
\nomenclature[B]{\(c_{i,i^{\prime}}^{CP}/c_{i,i^{\prime}}^{CF}\)}{Cost of inter-area transmission capacity for generation capability/ capacity for flexibility and resilience from area $i$ to area $i^{\prime}$}
\nomenclature[B]{\(E^0/E^{\max}\)}{Minimum and maximum storage capacity of ES}
\nomenclature[B]{\(H_{i/k}\)}{Inertia constant}
\nomenclature[B]{\(P_l^{\max}\)}{Maximum transmission capacity of the inter-area transmission line}
\nomenclature[B]{\(P_{i,t,\omega}^d\)}{Power demand}
\nomenclature[B]{\(\Delta P_{i,t,\omega}^d\)}{Fluctuation of power demand}
\nomenclature[B]{\(P_{w,t,\omega}^{wind}/P_{s,t,\omega}^{solar}\)}{Outputs of wind power unit and photovoltaic units}
\nomenclature[B]{\(\Delta P_{w,t,\omega}^{wind}/\Delta P_{s,t,\omega}^{solar}\)}{Fluctuation of wind power and photovoltaic units}
\nomenclature[B]{\(P_{i,i^{\prime}}^{LC}/P_{i,i^{\prime}}^{LF}\)}{Upper limits for capacity that can be provided to the other areas for generation capability / for flexibility and resilience}
\nomenclature[B]{\(P_k^{\Delta,\max}\)}{Maximum investment capacities}
\nomenclature[B]{\(P_k^{\Theta}\)}{Existing capacities}
\nomenclature[B]{\(T^B\)}{The time of the blackout}
\nomenclature[B]{\(VOLL\)}{Penalty for loss caused by insufficient provisioned capacity}
\nomenclature[B]{\(\alpha_{k}\)}{Coefficient of confidence capacity of units generation capability / the capacity for flexibility and resilience}
\nomenclature[B]{\(\tau_{i/k}\)}{Available time coefficient}
\nomenclature[B]{\(\rho_{k}\)}{Annualized investment cost per MW}
\nomenclature[B]{\(\sigma_{i/k}\)}{Ramping coefficient}
\nomenclature[B]{\(\lambda_{i, max}^{CP/CF/CM/CR}\)}{Upper limits for market clearing prices of generation capability/ ramping/ inertia/ recovery capability product}
\nomenclature[B]{\(\lambda_{i,t,\omega,max}^{EI}\)}{Upper limits for market clearing prices of energy market}
\nomenclature[B]{\(a_{i,t,w}^{EI}\)}{Intercept of inverse demand function for energy market}
\nomenclature[B]{\(b_{i,t,w}^{EI}\)}{Slope of inverse demand function for energy market}
\nomenclature[B]{\(\delta^{DC/C}\)}{Power charging and discharging efficiency rate of ES}
\nomenclature[C]{\(F\)}{Ramping capability}
\nomenclature[C]{\(M\)}{Inertia}
\nomenclature[C]{\(P\)}{Generation capability}
\nomenclature[C]{\(R\)}{Recovery energy supply}
\nomenclature[C]{\(p_{t,\omega}^{ploss}\)}{Loss caused by insufficient provisioned capacity for generation capability}
\nomenclature[C]{\(p_{t,\omega}^{floss}\)}{Loss caused by insufficient provisioned capacity for ramping capability}
\nomenclature[C]{\(m_{t,\omega}^{iloss}\)}{Loss caused by insufficient provisioned capacity for inertia}
\nomenclature[C]{\(e_{t,\omega}^{rloss}\)}{Loss caused by insufficient provisioned capacity for recovery capability}
\nomenclature[C]{\(P_{k}^{C}/P_{k}^{F}\)}{Provisioned capacity for generation capability/ capacity for flexibility and resilience}
\nomenclature[C]{\(P_{k}^{CI}/P_{k}^{FI}\)}{Schedule capacity for generation capability/ capacity for flexibility and resilience}
\nomenclature[C]{\(P_{i,i^{\prime}}^{CO}/P_{i,i^{\prime}}^{FO}\)}{Inter-area transmission capacity for generation capability/ capacity for flexibility and resilience from area $i$ to area $i^{\prime}$}
\nomenclature[C]{\(P_{l,t,w}^{EL}\)}{Power transmitted on the inter-area transmission line}
\nomenclature[C]{\(P_{k}^{\Delta}\)}{Investment capacity}
\nomenclature[C]{\(P_{k,t,\omega}^{E}\)}{Schedule capacity in the electricity market}
\nomenclature[C]{\(P_{j,e,t,\omega}^{EC/EDC}\)}{Power charging/discharging of ES}
\nomenclature[C]{\(E_{j,e,t,\omega}\)}{State of charge for ES}
\nomenclature[C]{\(\lambda_{i}^{CP/CF/CM/CR}\)}{Market clearing prices of generation capability product/ ramping capability product/ inertia product/ recovery capability product}
\nomenclature[C]{\(\lambda_{i,t,\omega}^{EI}\)}{Market clearing prices of energy market}
\nomenclature[C]{\(L_{i}^{CP/CF/CM/CR}\)}{Net injection to area $i$ of generation capability product/ ramping capability product/ inertia product/ recovery capability product}
\nomenclature[C]{\(a_{i,d}^{CP/CF/CM/CR}\)}{Intercept of inverse demand function for generation capability/ ramping/ inertia/ recovery capacity market}
\nomenclature[C]{\(b_{i,d}^{CP/CF/CM/CR}\)}{Slope of inverse demand function for generation capability/ ramping/ inertia/ recovery capacity market}
\nomenclature[C]{\(L_{i,t,w}^{EI}\)}{Net injection to  area $i$ in energy market}
\printnomenclature

\section{Introduction}

\IEEEPARstart{T}{he} integration of renewable energy (RE) into power systems has become a worldwide consensus in response to growing environmental challenges and energy security concerns. However, the rapidly increasing proportion of RE leads to the shortages of flexibility and resilience in power systems \cite{Guerra2022}. Such scarcity of flexibility and resilience has been witnessed in recent power outage incidents. For instance, South Australia has experienced a loss of 1,885 MW of load and a 50-hour blackout due to the lack of inertia and unsuccessful recovery \cite{A.E.M.Operator2016}.

As a result, several countries have modified the existing market mechanisms to satisfy the demand for flexibility and resilience for the secure operation of power systems. These modifications include revisions to existing frequency regulation \cite{Shiltz2016} and reserve \cite{Silva-Rodriguez2024} ancillary service markets, as well as introducing novel ancillary service products including the flexible ramping \cite{Cui2018} and inertia \cite{Qiu2024}. However, these market modifications ensure flexibility and resilience in real-time operation, assuming that the resources are adequate and available for real-time dispatch. In fact, as the proportion of RE rapidly increases and RE is typically unable to provide flexibility and resilience, the adequacy of resources for flexibility and resilience within the power systems could be doubted \cite{Panteli2017, Ma2013, Zhu2024}. 

As an effective market mechanism to stimulate investment in generation units and guarantee the adequacy of generation capability, the capacity market has been practiced in many countries, including the United States \cite{Bowring2013} and the United Kingdom \cite{Harbord2014}. The state-of-the-art studies on capacity market mainly focus on integrating the new types of generation resources, such as RE and energy storage (ES). Specifically, the discussions concentrate on how multiple types of resources including RE, ES and demand response participate in the capacity market \cite{Khan2018} and how to identify their crediting capacities \cite{Wang2024}. Formulating the capacity demand curve is a critical task in designing a capacity market, as an efficient capacity demand curve can enhance market efficiency, reduce operating costs, and ensure capacity adequacy \cite{Hobbs2007}. Conleigh et al. \cite{Byers2018} compared four ISOs' capacity demand curve design methodologies in the United States. Feng et al. \cite{Zhao2017} proposed a model aiming at minimizing the total cost of capacity and expected load loss, with capacity demand curves obtained by solving the model in conjunction with stochastic production simulation. Steffen et al. \cite{Kaminski2023} proposed a methodology for designing capacity demand curves that reflect investors' risk preferences and the costs of newly entered units. In addition to studies focusing on existing capacity market mechanisms, some research has incorporated the provision of flexibility and resilience into the capacity market. Fang et al. \cite{Fang2021} proposed a capacity market clearing model with constraints on flexible ramping demand. Hu et al. \cite{Hu2023} and Liang et al. \cite{Liang2023} addressed the shortage of inertia in power systems with high RE penetration, and proposed the inertia capacity markets as well as the corresponding pricing schemes to guarantee long-term inertia adequacy. Saraf et al. \cite{Saraf2009} analyzed the selections and awards for the units providing recovery capability in ERCOT. To summarize, these studies have incorporated flexibility or resilience demand into the capacity market. However, each study focuses on a single type of demand, lacking discussions on the mutual effects of different kinds of capacity demand, and most importantly, how these multiple types of demands collectively impact generation investment, which is significant as this oversight hinders the verification of whether the capacity market efficiently and properly stimulates generation investment and guarantees long-run resource adequacy.

To effectively stimulate generation investment, formulating the demand curve in the capacity market needs to consider the behavior of generation investment of different generation companies (GenCos). The generation investment model has been established in several studies. Grimm \cite{Grimm2021} established a multi-level Genco’s generation investment model under different market designs of price zones. Bhagwat \cite{Bhagwat2017} established the generation investment model of GenCos under the forward capacity market, yearly clearing capacity market and energy-only market, verifying that both types of capacity market could increase generation investment, while the forward capacity market increases more in generation investment of low-cost peak generation capacity. Hach \cite{Hach2016} proposed a generation investment model for Gencos under a capacity market dedicated to new investments, compared to a capacity market for both existing and newly-entered capacities, finding that the former leads to more generation investment. The generation investment among multiple Gencos creates an equilibrium problem, often addressed using game-theoretic approaches. Wogrin \cite{Wogrin2013} formed the generation investment equilibrium in the electricity and natural gas markets into an equilibrium problem with equilibrium constraints. Mousavian \cite{Mousavian2020} established an equilibrium model with the first stage of generation investment and the second stage of the spot market, identifying the consistent equilibrium through the conjectural variations approach. An ADMM-based method was applied to solve the generation investment equilibrium in the capacity market considering risk-averse agents in \cite{Hoschle2018}. Existing literature studies the generation investment equilibrium with given generation capacities, while rarely concentrating on formulating the capacity demand curve, considering its guidance on the generation investment equilibrium. To optimize the generation mix to enhance flexibility and resilience, generation investment and equilibrium should be considered when formulating the capacity demand curves. However, developing such a capacity demand curve considering the guidance on generation investment equilibrium would be more intricate than merely solving the generation investment equilibrium with predefined generation capacities, as this could lead to an equilibrium adjustment problem rather than a mere equilibrium solution. 

Moreover, in those countries or regions with multi-area coupled electricity markets, such as Europe and China \cite{Ihlemann2022, Li2022}, the design of capacity market should also consider its guidance on generation investment in each area as it has been pointed out that inappropriate capacity market mechanism in multi-area coupled electricity markets may harm producers’ profits \cite{Cepeda2018}, and decrease the capacity adequacy \cite{Meyer2015} as the generation mixes in each area are mutually affected.

This paper aims to incorporate the flexibility and resilience demand into the capacity market, considering the guidance on generation investment. The main contributions are summarized as follows:

1) We specify the flexibility and resilience demand in the capacity market as three trading products: ramping capability, inertia and recovery capability, and explore the corresponding capacity demand quantization models. 

2) We propose a capacity demand curve formulation model that considers the generation investment equilibrium among Gencos. A Nash-Cournot model with an equivalent quadratic programming formulation solves the equilibrium among Gencos. An iterative approach is designed to solve the tri-level model, which coordinates the capacity demand curve formulation and the generation investment equilibrium problem.

3) The case study validates that the proposed formulation of the capacity demand curve, which incorporates the demand for flexibility and resilience into the capacity market, could stimulate proper generation investment and ensure the adequacy of flexibility and resilience for power systems. The case study also justifies the feasibility and efficiency of the proposed methodology in interconnected systems.

The remainder of this paper is organized as follows: Section II analyzes and quantifies the capacity demand for ramping capability, inertia, and recovery capability and proposes a capacity market mechanism. Section III establishes the capacity demand curve formulation model considering generation investment equilibrium among Gencos. Section IV proposes the solving methods, and Section V provides the case study. Section VI concludes the work.

\section{Description of the flexibility and resilience demand and the capacity market framework}

\begin{figure}
  \begin{center}
  \includegraphics[width=3.5in]{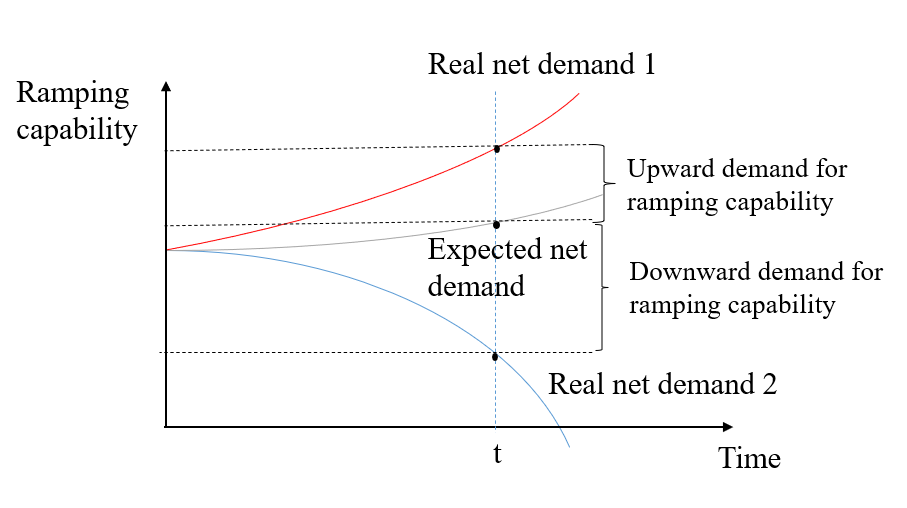}\\
  \caption{Schematic diagram of demand for ramping capability}\label{ramping capability}
  \end{center}
\end{figure}

Flexibility and resilience are among the main requirements for the stable operation of power systems with a high proportion of RE. Specifically, flexibility refers to the ability to track RE fluctuations quickly \cite{Brunner2020}. At the same time, resilience denotes the ability to recover from extreme events swiftly \cite{Bie2017}, which could happen more frequently due to the increased openness of the power systems with high RE penetration \cite{Bie2017}. To reflect the flexibility and resilience demand, we select the ramping capability, inertia, and recovery capability \cite{Bie2017, Makolo2021, Wang2017} as the novel trading products in the capacity market. This section discusses in detail the concept of these trading products and then proposes the capacity market framework.

\subsection {Ramping capability}

The ramping capability is referred to as the capability that could realize the quick power output increase or decrease to track the fluctuation of net demand, as depicted in Fig.1.

The overall demand for ramping capability of the power system can be expressed as: 
\begin{equation}\label{nonideal_rectifier_resistance}
{{F}_{t}}=\sum\limits_{i}{(\Delta P_{i,t}^{d})}-\sum\limits_{w\in {{\Omega }_{w}}}{(\Delta P_{w,t}^{wind})-}\sum\limits_{k\in {{\Omega }_{s}}}{(\Delta P_{s,t}^{solar})};\forall t
\end{equation}

The units’ capacities satisfying the flexibility and resilience demand could provide ramping capability based on their ramping performance, and the ramping capability that unit $k$ provides is represented as follows, where the ramping performance coefficient is referred to as the ramp rate ratio of unit $k$ times its maximum generation capacity \cite{Fang2021}:

\begin{equation}\label{nonideal_rectifier_resistance}
F_{k}^{{}}={{\sigma }_{k}}P_{k}^{FI};k\notin (\Omega _{i}^{S}\cup \Omega _{i}^{W})
\end{equation}

\subsection {Inertia}

The inertia indicates the ability to store kinetic energy of the synchronous machines connected to the power grid. RE's incapability to provide inertia results in frequency fluctuations, leading to a rapidly increasing demand for inertia. Power systems with insufficient inertia typically exhibit prominent frequency issues, such as the increase in the rate of change of frequency (RoCoF) and a downward shift in the lowest frequency point ($f_{nadir}$) of the power system, as illustrated in Fig.2.

\begin{figure}
  \begin{center}
  \includegraphics[width=2.75in]{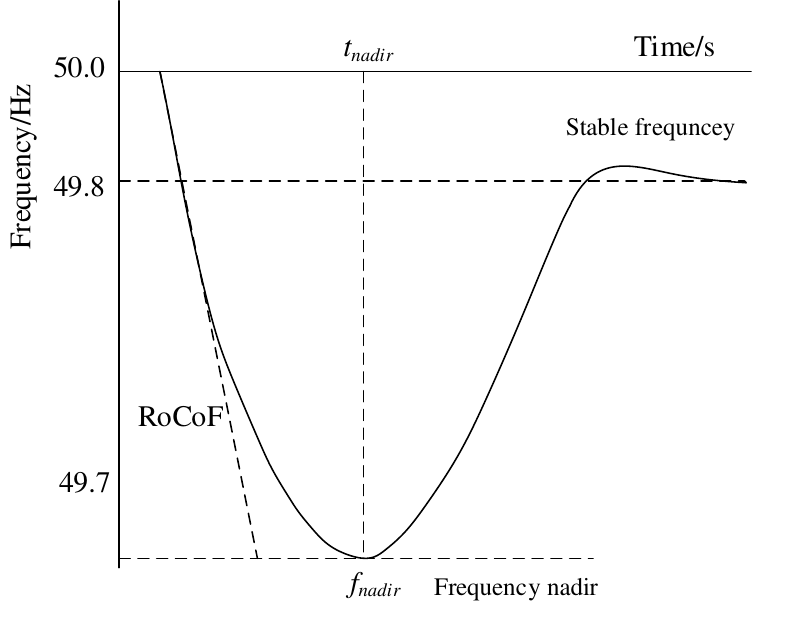}\\
  \caption{Schematic diagram of demand for inertia}\label{circuit_diagram}
  \end{center}
\end{figure}

To ensure that the RoCoF and the lowest frequency at nadir meet the requirements for the stable operation of the power system, the corresponding inertia requirements are represented as follows:

\begin{equation}\label{diode_voltage_waveform_B}
 \begin{aligned}
  & {{M}_{t}}=\max \{{}^{{{f}_{0}}P_{t}^{loss}}/{}_{2RoCo{{F}_{\max }}}, \\ 
 & T(Rg-Fg)(\frac{P_{t}^{loss}{{e}^{-\zeta \delta \upsilon }}}{\Delta {{f}_{\max }}(D+Rg)+P_{t,\omega }^{loss}})\};\forall t \\ 
 &  \\ 
\end{aligned}
\end{equation}

where $\Delta {{f}_{\max }}$is the maximum admissible frequency deviation at nadir. $T$ is the generator's time constant. $\upsilon$ is the time from the start of the accident to the frequency nadir. $Rg$, $Fg$, $D$, $\varsigma$ and $\delta$ represent the droop coefficients, the proportion of power generated by the synchronous generator, the damping constant, the damping ratio, and the natural frequency, respectively.

The first term in (3) represents the inertia demand determined by the RoCoF, while the second term represents the demand determined by the frequency nadir. During the real-world operation of the power system, the problems caused by RoCoF exceeding the limit are more common and severe than those caused by deviations in the frequency at nadir \cite{Liang2023}. Therefore, this work adopts the inertia demand determined by the RoCoF limit as the actual inertia demand in the capacity market.

The units’ capacities satisfying the flexibility and resilience demand could provide inertia based on their inertia constant. Thus, the inertia that unit $k$ provides is represented as

\begin{equation}\label{diode_current_waveform_time_domain}
   M_{k}^{{}}={{H}_{k}}P_{k}^{FI};k\notin (\Omega _{i}^{S}\cup \Omega _{i}^{W})
\end{equation}

\subsection {Recovery capability}

Due to the impact of grid-connected REs, electronic devices, hybrid AC/DC, inter-area power system interconnection and other complex operating modes in power systems \cite{Yang2022}, blackouts are becoming increasingly frequent. Unsuccessful restoration from blackouts in a short time would cause severe economic losses. The process of a unit providing recovery energy supply is shown in Fig. 3. Since a unit usually requires a specific period to start and ramp up to its rated power output, a certain amount of capacity is required to quickly provide the recovery power supply and minimize the economic losses of the load during the blackout, which is referred to as the recovery capability.

\begin{figure}
  \begin{center}
  \includegraphics[width=3.5in]{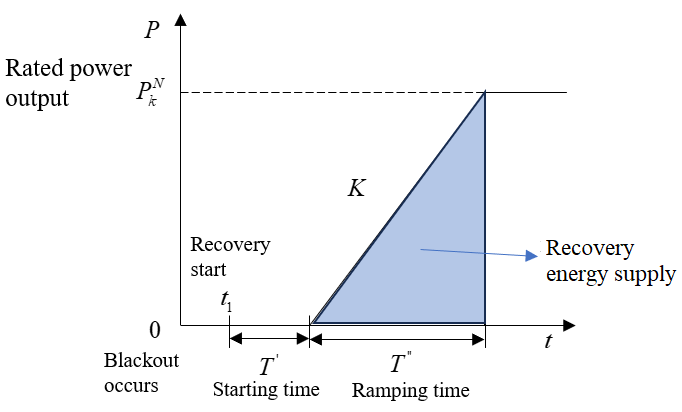}\\
  \caption{Process of a unit to provide recovery energy supply}\label{circuit_diagram}
  \end{center}
\end{figure}
During the recovery process, the energy loss of loads can be represented as:

\begin{equation}\label{diode_current_waveform_time_domain}
   {{R}_{t}}=\sum\limits_{i}{\partial _{i}^{load}\pi _{i}^{loss}(P_{i,t}^{d}){{T}^{B}}};\forall t,\omega 
\end{equation}

The units’ capacities satisfying the flexibility and resilience demand could provide recovery capability. Still, according to Fig.3, it may not offer recovery energy supply at its capacity for recovery capability for the whole recovery process. Therefore, the available time coefficient of the capacity, which considers the unit's startup time and ramping performance, is selected to reflect the relationship between the recovery energy supply and the capacity for recovery capability. The recovery energy supply provided by unit $k$ is described as follows:

\begin{equation}\label{diode_current_waveform_time_domain}
 R_{k}^{{}}={{\tau }_{k}}P_{k}^{FI};k\notin (\Omega _{i}^{S}\cup \Omega _{i}^{W})
\end{equation}

\subsection {Capacity market framework}

Fig. 4 illustrates the capacity market framework. The capacity market with flexibility and resilience demand contains four trading products in total: the generation capability (identical to the existing capacity market), as well as the ramping capability, inertia, and recovery capability, as introduced in this work.

\begin{figure}
  \begin{center}
  \includegraphics[width=3.5in]{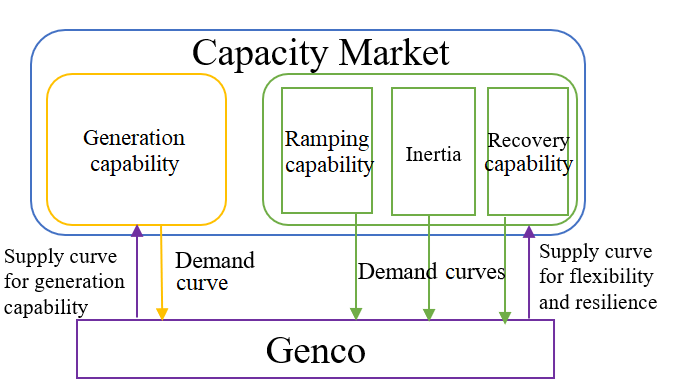}\\
  \caption{Schematic diagram of the capacity market framework}\label{circuit_diagram}
  \end{center}
\end{figure}

 The capacity demand curves are formulated and released for each trading product. At the same time, market participants submit two supply curves, one for generation capability and the other for the rest of the three products, also referred to as the supply curve for flexibility and resilience. The reason for formulating a single supply curve for the three trading products can be specified as follows. The ramping capability, inertia, and recovery capability can be categorized as ancillary services. The capacities for generation capability should be reserved for providing generation supply and cannot be dispatched to provide ancillary services, while various types of ancillary services have a certain degree of substitutability \cite{Chicco2004}. Thus, the provisioned capacity reserved for providing flexibility and resilience could exhibit ramping capability, inertia and recovery capability based on their ramping performance, inertia constant and the available time coefficient, implied by (2), (4) and (6). The trading product of generation capability is cleared using the demand and supply curves for generation capability; then, the provisioned capacity for generation capability and clearing prices are determined. The trading products of ramping capability, inertia and recovery capability are cleared using the capacity demand curves for each of these three trading products, along with a single supply curve for flexibility and resilience. This process determines three clearing prices (each for ramping capability, inertia, and recovery capability) and one provisioned capacity for flexibility and resilience.

\section {Mathematical models for formulating capacity demand curves considering the guidance on generation investment}

This section describes the framework and mathematical models for formulating the capacity demand curves, considering the guidance on generation investment. The problem is structured as a tri-level model, demonstrating the interactions among the demand curve formulator (responsible for formulating capacity demand curves, often acted by the market operator \cite{PJM2022}), GenCos and the market operator. In the upper level, the capacity demand curves are formulated for the four trading products—generation capability, ramping capability, inertia, and recovery capability—based on the generation investment capacities of GenCos, and the inter-area transmission capacity obtained from the middle and lower levels. Identical to the existing capacity markets \cite{Byers2018}, the demand curves are formulated in a decreasing, piecewise-linear form, discretized into segments through a series of inflection points, with capacity demand as the horizontal axis and price as the vertical axis, indicated as the red line in Fig.5.

\begin{figure}
  \begin{center}
  \includegraphics[width=3.5in]{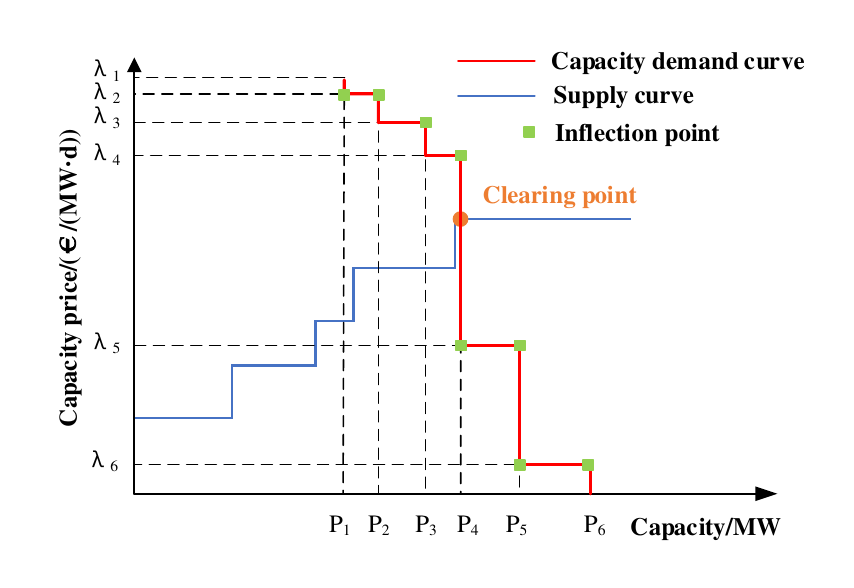}\\
  \caption{Schematic diagram of capacity demand curve}\label{circuit_diagram}
  \end{center}
\end{figure}

In the middle level, GenCos determine their generation investment strategies in the capacity and energy markets based on the demand curves and the market-clearing prices. The lower-level models facilitate the market clearing of capacity and energy markets. Fig. 6 shows the schematic diagram of the tri-level model.

\begin{figure*}
  \begin{center}
  \includegraphics[width=4in]{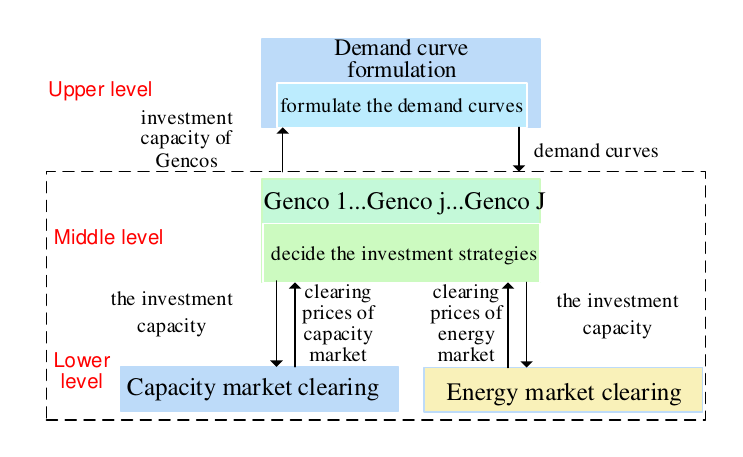}\\
  \caption{Schematic diagram of the model for formulating capacity demand curves considering the guidance on generation investment}\label{circuit_diagram}
  \end{center}
\end{figure*}

\subsection {Upper level: the capacity demand curve formulation model}
The mathematical model for formulating the capacity demand curves is illustrated as follows:

\begin{equation}\label{fundamental_diode_voltage}
 \min {{C}^{C}}+{{C}^{F}}+CPU{{E}_{\omega }}+CFU{{E}_{\omega }}
\end{equation}
\begin{equation}\label{fundamental_diode_voltage}
{{C}^{C}}=\sum\limits_{k\in \Omega _{i}^{K}}{c_{k}^{CP}}P_{k}^{C}+\sum\limits_{i\in \Omega _{i}^{i^{\prime}}}{c_{i}^{CP}P_{i,i^{\prime}}^{CO}}
\end{equation}
\begin{equation}\label{fundamental_diode_voltage}
{{C}^{F}}=\sum\limits_{k\in \Omega _{i}^{K}}{c_{k}^{CF}}P_{k}^{F}+\sum\limits_{i\in \Omega _{i}^{i^{\prime}}}{c_{i}^{CF}P_{i,i^{\prime}}^{FO}}
\end{equation}
\begin{equation}\label{fundamental_diode_voltage}
CPU{{E}_{\omega }}=VOL{{L}_{peak}}*p_{t,\omega }^{ploss};\forall t,\omega 
\end{equation}
\begin{equation}\label{fundamental_diode_voltage}
 \begin{aligned}
  & CFU{{E}_{\omega }}=VOL{{L}_{fluctaute}}*p_{t,\omega }^{floss}+VOL{{L}_{inertia}}*m_{t,\omega }^{iloss} \\ 
 & +VOL{{L}_{recover}}*e_{t,\omega }^{rloss};\forall t,\omega  \\ 
\end{aligned}
\end{equation}
\begin{equation}\label{fundamental_diode_voltage}
 \begin{aligned}
&p_{t,\omega }^{ploss}= \\
&{{[\sum\limits_{i}{(P_{i,t,\omega }^{d})}-\sum\limits_{w\in \Omega _{i}^{W}}{(P_{w,t,\omega }^{wind})-}\sum\limits_{s\in \Omega _{i}^{S}}{(P_{s,t,\omega }^{solar})}-P_{i,q}^{{}}]}^{+}};\forall i,t,\omega 
\end{aligned}
\end{equation}
\begin{equation}\label{fundamental_diode_voltage}
\begin{aligned}
  & p_{t,\omega }^{floss}=\\
  &\max \{{{[\sum\limits_{i}{(\Delta P_{i,t,\omega }^{d})}-\sum\limits_{w\in \Omega _{i}^{W}}{(\Delta P_{w,t,\omega }^{wind})-}\sum\limits_{s\in \Omega _{i}^{S}}{(\Delta P_{s,t,\omega }^{solar})}-F_{i,q}^{{}}]}^{+}}, \\
 & {{[\sum\limits_{i}{(\Delta P_{i,t,\omega }^{d})}-\sum\limits_{w\in \Omega _{i}^{W}}{(\Delta P_{w,t,\omega }^{wind})-}\sum\limits_{s\in \Omega _{i}^{S}}{(\Delta P_{s,t,\omega }^{solar})}-(-F_{i,q}^{{}})]}^{+}}\};\\
 &\forall i,t,\omega  \\ 
\end{aligned}
\end{equation}
\begin{equation}\label{fundamental_diode_voltage}
\begin{aligned}
  & m_{t,\omega }^{iloss}={{[{}^{{{f}_{0}}P_{t,\omega }^{loss}}/{}_{2RoCo{{F}_{\max }}}-M_{i,q}^{{}}]}^{+}} \\ 
 & P_{t,\omega }^{loss}=\\
& {{(\sum\limits_{i}{(\Delta P_{i,t,\omega }^{d})}-\sum\limits_{w\in \Omega _{i}^{W}}{(\Delta P_{w,t,\omega }^{wind})-}\sum\limits_{s\in \Omega _{i}^{S}}{(\Delta P_{s,t,\omega }^{solar})})}^{+}};\forall i,t,\omega  \\ 
\end{aligned}
\end{equation}
\begin{equation}\label{fundamental_diode_voltage}
 e_{t,\omega }^{rloss}={{[\sum\limits_{i}{\partial _{i}^{load}\pi _{i}^{loss}(P_{i,t,\omega }^{d}){{T}^{B}}}-R_{i,q}^{{}}]}^{+}};\forall i,t,\omega 
\end{equation}
\begin{equation}\label{fundamental_diode_voltage}
P_{i,q}^{{}}=\sum\limits_{k\in \Omega _{i}^{K}}{P_{k}^{C}}+\sum\limits_{k\in \Omega _{i}^{K}}^{k\notin (\Omega _{i}^{S}\cup \Omega _{i}^{W})}{P_{k}^{F}+\sum\limits_{i\in \Omega _{i}^{i^{\prime}}}{P_{i,i^{\prime}}^{CO}}}+\sum\limits_{i\in \Omega _{i}^{i^{\prime}}}{P_{i,i^{\prime}}^{FO}};\forall i
\end{equation}
\begin{equation}\label{fundamental_diode_voltage}
F_{i,q}^{{}}=\sum\limits_{k\in \Omega _{i}^{K}}^{k\notin (\Omega _{i}^{S}\cup \Omega _{i}^{W})}{{{\sigma }_{k}}P_{k}^{F}}+\sum\limits_{i\in \Omega _{i}^{i^{\prime}}}{{{\sigma }_{i}}P_{i,i^{\prime}}^{FO}};\forall i
\end{equation}
\begin{equation}\label{fundamental_diode_voltage}
 M_{i,q}^{{}}=\sum\limits_{k\in \Omega _{i}^{K}}^{k\notin (\Omega _{i}^{S}\cup \Omega _{i}^{W})}{{{H}_{k}}P_{k}^{F}+\sum\limits_{i\in \Omega _{i}^{i^{\prime}}}{{{H}_{i}}P_{i,i^{\prime}}^{FO}}};\forall i
\end{equation}
\begin{equation}\label{fundamental_diode_voltage}
R_{i,q}^{{}}=\sum\limits_{k\in \Omega _{i}^{K}}^{k\notin (\Omega _{i}^{S}\cup \Omega _{i}^{W})}{{{\tau }_{k}}P_{k}^{F}}+\sum\limits_{i\in \Omega _{i}^{i^{\prime}}}{{{\tau }_{i}}P_{i,i^{\prime}}^{FO}};\forall i
\end{equation}
\begin{equation}\label{fundamental_diode_voltage}
P_{k}^{C}\le {{\alpha }_{k}}(P_{k}^{\Delta }+P_{k}^{\Theta });\forall i,k
\end{equation}
\begin{equation}\label{fundamental_diode_voltage}
P_{k}^{C}+P_{k}^{F}\le {{\alpha }_{k}}(P_{k}^{\Delta }+P_{k}^{\Theta });\forall i,k\notin ({{\Omega }_{s}}\cup {{\Omega }_{w}})
\end{equation}
\begin{equation}\label{fundamental_diode_voltage}
P_{k}^{C},P_{k}^{F}\ge 0;\forall i,k
\end{equation}
\begin{equation}\label{fundamental_diode_voltage}
-P_{i,i^{\prime}}^{LC}\le P_{i,i^{\prime}}^{CO}\le P_{i,i^{\prime}}^{LC};\forall i
\end{equation}
\begin{equation}\label{fundamental_diode_voltage}
-P_{i,i^{\prime}}^{LF}\le P_{i,i^{\prime}}^{FO}\le P_{i,i^{\prime}}^{LF};\forall i
\end{equation}
\begin{equation}\label{fundamental_diode_voltage}
-P_{l}^{\max }\le P_{i,i^{\prime}}^{CO}+P_{i,i^{\prime}}^{FO}\le P_{l}^{\max };\forall i
\end{equation}

The objective function (7) represents the total costs, which contain the costs for purchasing capacity and the expected economic losses caused by insufficient provisioned capacity when formulating the demand curves. (8) and (9) represent the expected cost for purchasing capacity, while (10) and (11) represent the expected economic losses of insufficient provisioned capacity. The loss of load caused by insufficient provisioned capacities of generation capability, ramping capability, inertia and recovery capability is calculated in (12) - (15), respectively. (16) - (19) represent the constraints of the overall provisioned capacities for generation, ramping, inertia, and recovery capability, respectively. (20) - (22) set the limits for the provisioned capacities of units in the capacity market. RE can only provide the generation capability product. In contrast, other types of units, such as thermal units (TUs) and ESs, can provide products of generation, ramping, inertia, and recovery capability. (23) - (25) impose constraints on inter-area transmission capacity. It should be noted that in the above formulas, the subscript q represents the inflection points on the capacity demand curves. By calculating the expected economic losses caused by insufficient provisioned capacity and the expected cost for purchasing capacity at a series of inflection points of capacity demand curves, the capacity demand curves can be acquired by obtaining the corresponding prices of each inflection point by the dual multipliers of (16) - (19). 

\subsection {Middle level: Genco’s generation investment model}
a. Genco’s decision model

To pursue higher profits, each Genco $j$ decides on their generation investment strategies in both capacity and energy markets, and the model is illustrated as follows: The decision variables are $\left\{ P_{j,k}^{\Delta }, P_{j,k}^{CI}, P_{j,k}^{FI}, P_{j,k}^{CI}, P_{j,k}^{FI}, P_{j,k,t,\omega }^{E}, P_{j,e,t,\omega }^{EC}, P_{j,e,t,\omega }^{EDC} \right\}$.

\begin{equation}\label{fundamental_diode_voltage}
\begin{matrix}
  \min \sum\limits_{k\in {{\Omega }_{Ka}}}{{{\rho }_{k}}P_{j,k}^{\Delta }}+(8760/T)*\sum\limits_{k\in {{\Omega }_{Ka}}}{\sum\limits_{\omega }{\sum\limits_{t=1}^{T}{{{\pi }_{\omega }}{{C}_{k}}(P_{j,k,t,\omega }^{E})}}} \\ 
   -(8760/T)*\sum\limits_{k\in {{\Omega }_{Ka}}}{\sum\limits_{\omega }{\sum\limits_{t=1}^{T}{{{\pi }_{\omega }}(\lambda _{i,t,\omega }^{EI}P_{j,k,t,\omega }^{E})}}} \\
  -365*\sum\limits_{k\in {{\Omega }_{Ka}}}{(\lambda _{i}^{CP}(P_{j,k}^{CI}+P_{j,k}^{FI})+(\lambda _{i}^{CF}+\lambda _{i}^{CM}+\lambda _{i}^{CR})P_{j,k}^{FI}}) \\ 
\end{matrix}
\end{equation}
\begin{equation}\label{fundamental_diode_voltage}
0\le P_{j,k,t,\omega }^{E}\le P_{j,k}^{\Delta }+P_{j,k}^{\Theta }
\end{equation}
\begin{equation}\label{fundamental_diode_voltage}
0\le P_{j,e,t,\omega }^{EC}\le P_{j,e}^{\Delta }+P_{j,e}^{\Theta }
\end{equation}
\begin{equation}\label{fundamental_diode_voltage}
0\le P_{j,e,t,\omega }^{EDC}\le P_{j,e}^{\Delta }+P_{j,e}^{\Theta }
\end{equation}
\begin{equation}\label{fundamental_diode_voltage}
P_{j,k}^{CI}\le {{\alpha }_{j,k}}(P_{j,k}^{\Delta }+P_{j,k}^{\Theta });\forall j, k
\end{equation}
\begin{equation}\label{fundamental_diode_voltage}
P_{j,k}^{CI}+P_{j,k}^{FI}\le {{\alpha }_{j,k}}(P_{j,k}^{\Delta }+P_{j,k}^{\Theta });k\notin ({{\Omega }_{s}}\cup {{\Omega }_{w}});\forall j, k
\end{equation}
\begin{equation}\label{fundamental_diode_voltage}
P_{j,k}^{CI},P_{j,k}^{FI}\ge 0;\forall j,k
\end{equation}
\begin{equation}\label{fundamental_diode_voltage}
0\le P_{j,k}^{\Delta }\le P_{j,k}^{\Delta ,\max };\forall j, k\in {{\Omega }_{Ka}}
\end{equation}
\begin{equation}\label{fundamental_diode_voltage}
0\le \sum\limits_{k\in {{\Omega }_{Ka}}}{P_{j,k}^{\Delta }}\le {{\partial }_{Ka}}\sum\limits_{k\in {{\Omega }_{Ka}}}{P_{j,k}^{\Theta }}
\end{equation}

The objective function (26) represents that GenCos consider the investment costs, power generation costs, the total revenue in the capacity market, and the total revenue in the energy market when deciding on their generation investment. (30) - (32) set limits on Gencos's provisioned capacities for generation capability, flexibility, and resilience. (33) and (34) impose investment constraints, including the maximum investment limits for each GenCos.

b. Transco’s decision model

\begin{equation}\label{fundamental_diode_voltage}
-P_{i,i^{\prime}}^{\max }\le P_{i,i^{\prime}}^{CO}\le P_{i,i^{\prime}}^{\max };\forall i,i^{\prime}
\end{equation}
\begin{equation}\label{fundamental_diode_voltage}
-P_{i,i^{\prime}}^{\max }\le P_{i,i^{\prime}}^{FO}\le P_{i,i^{\prime}}^{\max };\forall i,i^{\prime}
\end{equation}
\begin{equation}\label{fundamental_diode_voltage}
-P_{i,i^{\prime}}^{\max }\le P_{i,i^{\prime}}^{CO}+P_{i,i^{\prime}}^{FO}\le P_{i,i^{\prime}}^{\max };\forall i,,i^{\prime}
\end{equation}
\begin{equation}\label{fundamental_diode_voltage}
-P_{l}^{\max }\le P_{l,t,\omega }^{EL}\le P_{l}^{\max }; \forall l
\end{equation}

(35)—(38) are the limits for inter-area transmission capacity, and (35)—(37) limit the inter-area transmission capacity in the capacity market. (38) limits the energy market's power flow through transmission lines.
\subsection{Lower level: market clearing models}

a. The capacity market model

\begin{equation}\label{fundamental_diode_voltage}
\begin{matrix}
   \lambda _{i,d}^{CP}=a_{i,d}^{CP}-b_{i,d}^{CP}\left( L_{i}^{CP} \right)  \\
   \lambda _{i,d}^{CF}=a_{i,d}^{CF}-b_{i,d}^{CF}\left( L_{i}^{CF} \right)  \\
   \lambda _{i,d}^{CM}=a_{i,d}^{CM}-b_{i,d}^{CM}\left( L_{i}^{CM} \right)  \\
   \lambda _{i,d}^{CR}=a_{i,d}^{CR}-b_{i,d}^{CR}\left( L_{i}^{CR} \right);\forall i\in I  \\
\end{matrix}
\end{equation}

(39) is the inverse demand function of the capacity market, which defines the zonal capacity energy price as a linear function of the total zonal sales, with $\lambda _{i,d}^{CP}$, $\lambda _{i,d}^{CF}$, $\lambda _{i,d}^{CM}$, $\lambda _{i,d}^{CR}$ being the clearing prices for the corresponding generation capability, ramping, inertia, and recovery capability market products.

\begin{equation}\label{fundamental_diode_voltage}
\begin{aligned}
  & 0\le \lambda _{i}^{CP}\le \lambda _{i,\max }^{CP}, \\ 
 & 0\le \lambda _{i}^{CF}\le \lambda _{i,\max }^{CF}, \\ 
 & 0\le \lambda _{i}^{CM}\le \lambda _{i,\max }^{CM}, \\ 
 & 0\le \lambda _{i}^{CR}\le \lambda _{i,\max }^{CR}; \forall i\in I \\ 
\end{aligned}
\end{equation}

(40) sets limits on the market-clearing prices.

\begin{equation}\label{fundamental_diode_voltage}
\sum\limits_{k\in \Omega _{i}^{K}}{P_{j,k,i}^{CI}}+\sum\limits_{k\in \Omega _{i}^{K}}{P_{j,k,i}^{FI}}+\sum\limits_{i\in \Omega _{i}^{i^{\prime}}}{P_{i,i^{\prime}}^{CO}}+\sum\limits_{i\in \Omega _{i}^{i^{\prime}}}{P_{i,i^{\prime}}^{COF}}=L_{i}^{CP};\forall i
\end{equation}
\begin{equation}\label{fundamental_diode_voltage}
\sum\limits_{k\in \Omega _{i}^{K}}^{k\notin (\Omega _{i}^{S}\cup \Omega _{i}^{W})}{{{\sigma }_{k}}P_{j,k,i}^{FI}}+\sum\limits_{i\in \Omega _{i}^{i^{\prime}}}{{{\sigma }_{i}}P_{i,i^{\prime}}^{FO}}=L_{i}^{CF};\forall i
\end{equation}
\begin{equation}\label{fundamental_diode_voltage}
\sum\limits_{k\in \Omega _{i}^{K}}^{k\notin (\Omega _{i}^{S}\cup \Omega _{i}^{W})}{{{H}_{k}}P_{j,k,i}^{FI}}+\sum\limits_{i\in \Omega _{i}^{i^{\prime}}}{{{H}_{i}}P_{i,i^{\prime}}^{FO}}=L_{i}^{CM};\forall i
\end{equation}
\begin{equation}\label{fundamental_diode_voltage}
\sum\limits_{k\in \Omega _{i}^{K}}^{k\notin (\Omega _{i}^{S}\cup \Omega _{i}^{W})}{{{\tau }_{k}}P_{j,k,i}^{FI}}+\sum\limits_{i\in \Omega _{i}^{i^{\prime}}}{{{\tau }_{i}}P_{i,i^{\prime}}^{FO}}=L_{^{i}}^{CR};\forall i
\end{equation}
\begin{equation}\label{fundamental_diode_voltage}
\sum\limits_{i}{P_{j,k,i}^{CI}}=P_{j,k}^{CI};\forall j,k
\end{equation}
\begin{equation}\label{fundamental_diode_voltage}
\sum\limits_{i}{P_{j,k,i}^{FI}}=P_{j,k}^{FI};\forall j,k
\end{equation}

(41)-(46) represent the supply-demand balance constraints of the capacities of generation capability, ramping capability, inertia, and recovery capability, respectively. It should be noted that provisioned capacities for flexibility and resilience can provide ramping, inertia, and recovery capability and contribute to generation capability. (45)-(46) calculate each Genco $j$'s total generation from different zone $i$ in the capacity market.

b. The energy market model

\begin{equation}\label{fundamental_diode_voltage}
\lambda _{i,t,w}^{EI}=a_{i,t,w}^{EI}-b_{i,t,w}^{EI}L_{i,t,w}^{EI};\forall i,t,w
\end{equation}
\begin{equation}\label{fundamental_diode_voltage}
0\le \lambda _{i,t,w}^{EI}\le \lambda _{i,t,w,\max }^{EI};\forall i,t,w
\end{equation}

(47) is the inverse demand function of the energy market. (48) set limits on the market-clearing prices.
\begin{equation}\label{fundamental_diode_voltage}
\begin{aligned}
  & \sum\limits_{w\in \Omega _{i}^{W}}{P_{j,i,w,t,\omega }^{wind}}+\sum\limits_{s\in \Omega _{i}^{S}}{P_{j,i,s,t,\omega }^{solar}}+\sum\limits_{k\in \Omega _{i}^{K}}^{k\notin (\Omega _{i}^{S}\cup \Omega _{i}^{W})}{P_{j,i,k,t,\omega }^{E}} \\ 
 & +\sum\limits_{e\in \Omega _{i}^{E}}{P_{j,i,e,t,\omega }^{EDC}}-\sum\limits_{e\in \Omega _{i}^{E}}{P_{j,i,e,t,\omega }^{EC}}+\sum\limits_{l\in \Omega _{i}^{L}}{{{A}_{l-i}}P_{l,t,\omega }^{EL}}=L_{i,t,w}^{EI}; \\ 
\end{aligned}
\end{equation}
\begin{equation}\label{fundamental_diode_voltage}
E_{j,e,t,\omega }^{{}}=E_{j,e,t-1,\omega }^{{}}-\frac{1}{{{\delta }^{DC}}}P_{j,e,t,\omega }^{EDC}+{{\delta }^{C}}P_{j,e,t,\omega }^{EC},\forall t>1
\end{equation}
\begin{equation}\label{fundamental_diode_voltage}
E_{j,e,T,\omega }^{{}}=E_{j,e,1,\omega }^{0}
\end{equation}
\begin{equation}\label{fundamental_diode_voltage}
E_{j,e,t,\omega }^{{}}\le {{E}^{\max }},\forall t>1
\end{equation}
\begin{equation}\label{fundamental_diode_voltage}
\sum\limits_{i}{P_{j,k,t,\omega,i }^{E}}=P_{j,k,t,\omega }^{E}; \forall j,t,w
\end{equation}
\begin{equation}\label{fundamental_diode_voltage}
\sum\limits_{i}{P_{j,e,t,\omega,i }^{EDC}}=P_{j,e,t,\omega }^{EDC}; \forall j,t,w
\end{equation}
\begin{equation}\label{fundamental_diode_voltage}
\sum\limits_{i}{P_{j,e,t,\omega,i }^{EC}}=P_{j,e,t,\omega }^{EC}; \forall j,t,w
\end{equation}

(49) represents the supply-demand balance constraint of the energy market. (50)-(52) indicate the constraints for ESs. (53)-(55) calculate each Genco $j$'s total generation from different zone $i$ in the energy market.  Specifically, the inverse demand function of the energy market, Eq. (47), is calibrated by calculating the coefficients using the generators’ expected long-run marginal cost \cite{S.Oliveira2023}. 

\section{The model reformulation and solving methodology}
\subsection {The reformulation for the middle and Lower levels Genco's investment problem model}

\begin{figure*}[!t]
\normalsize
\setcounter{MYtempeqncnt}{\value{equation}}
\setcounter{equation}{55}
\begin{equation}\label{fundamental_diode_voltage}
\begin{matrix}
  \max -\sum\limits_{j,k\in {{\Omega }_{Ka}}}{{{\rho }_{k}}P_{j,k}^{\Delta }}-(8760/T)*\sum\limits_{j,k\in {{\Omega }_{Ka}},{\omega },t}{{{\pi }_{\omega }}{{C}_{k}}(P_{j,k,t,\omega }^{E})} +(8760/T)*\sum\limits_{{\omega },i,t}{{{\pi }_{\omega }}\left( \left( a_{i,t,w}^{EI}-\frac{1}{2}b_{i,t,w}^{EI}L_{i,t,w}^{EI} \right)L_{i,t,\omega }^{EI} \right)} \\ 
  -(8760/T)*\sum\limits_{j,{\omega },i,t}{{{\pi }_{\omega }}
  \left( \frac{1}{2}b_{i,t,w}^{EI}{{\left(  \sum\limits_{w\in \Omega _{i}^{W}}{P_{i,w,t,\omega }^{wind}}+\sum\limits_{s\in \Omega _{i}^{S}}{P_{i,s,t,\omega }^{solar}} +\sum\limits_{k\in \Omega _{i}^{K}}^{k\notin (\Omega _{i}^{S}\cup \Omega _{i}^{W})}{P_{i,k,t,\omega }^{E}} +\sum\limits_{e\in \Omega _{i}^{E}}{P_{i,e,t,\omega }^{EDC}}-\sum\limits_{e\in \Omega _{i}^{E}}{P_{i,e,t,\omega }^{EC}} \right)}^{2}} \right)} \\ 
  +365*\sum\limits_{d,i}{\left( \left( a_{i,d}^{CP}-\frac{1}{2}b_{i,d}^{CP}\left( L_{i,d}^{CP} \right) \right)L_{i,d}^{CP} \right)} 
  -365*\sum\limits_{d,i}{\left( \frac{1}{2}b_{i,d}^{CP}{{\left( \sum\limits_{k\in \Omega _{i}^{K}}{P_{i,k}^{CI}} \right)}^{2}} \right)} \\ 
  +365*\sum\limits_{d,i}{\left( \left( a_{i,d}^{CF}-\frac{1}{2}b_{i,d}^{CF}\left( L_{i,d}^{CF} \right) \right)L_{i,d}^{CF} \right)} 
  -365*\sum\limits_{d,i}{\left( \frac{1}{2}b_{i,d}^{CF}{{\left( \sum\limits_{k\in \Omega _{i}^{K}}^{k\notin (\Omega _{i}^{S}\cup \Omega _{i}^{W})}{{{\sigma }_{k}}P_{i,k}^{FI}} \right)}^{2}} \right)} \\ 
  +365*\sum\limits_{d,i}{\left( \left( a_{i,d}^{CM}-\frac{1}{2}b_{i,d}^{CM}\left( L_{i,d}^{CM} \right) \right)L_{i,d}^{CM} \right)}  
  -365*\sum\limits_{d,i}{\left( \frac{1}{2}b_{i,d}^{CM}{{\left( \sum\limits_{k\in \Omega _{i}^{K}}^{k\notin (\Omega _{i}^{S}\cup \Omega _{i}^{W})}{{{H}_{k}}P_{i,k}^{FI}} \right)}^{2}} \right)} \\ 
  +365*\sum\limits_{d,i}{\left( \left( a_{i,d}^{CR}-\frac{1}{2}b_{i,d}^{CR}\left( L_{i,d}^{CR} \right) \right)L_{i,d}^{CR} \right)} 
  -365*\sum\limits_{d,i}{\left( \frac{1}{2}b_{i,d}^{CR}{{\left( \sum\limits_{k\in \Omega _{i}^{K}}^{k\notin (\Omega _{i}^{S}\cup \Omega _{i}^{W})}{{{\tau }_{k}}P_{i,k}^{FI}} \right)}^{2}} \right)} 
\end{matrix}
\end{equation}	

\begin{equation}\label{fundamental_diode_voltage}
(27-55)
\end{equation}
\setcounter{equation}{\value{MYtempeqncnt}}
\hrulefill
\vspace*{4pt}
\end{figure*}

As indicated in the above models, at the middle level, several GenCos decide on their generation investment and compete with each other in the lower-level energy and capacity market clearing problems. Therefore, the middle- and lower-level issues are equivalent to a Cournot-Nash model, which can be mathematically represented using Karush-Kuhn-Tucker (KKT) optimality conditions of each player's profit maximization problem, together forming a mixed-complementarity problem (MCP). In this paper, the mathematical property of the proposed model allows the formulation of an equivalent single optimization problem with a quadratic objective function to find the Cournot-Nash equilibrium. The solution and KKT conditions of such a problem are equivalent to those obtained with MCP formulations \cite{Hashimoto1985, Sohl1985, Hogan1997, Chattopadhyay2004} and might be solved with a wide range of NLP solvers such as GUROBI. 

Hence, the developed generation investment problem Objective:(26), Constraints: (27)-(55) can be reformulated in the centralized quadratic programming form as model Objective:(56), Constraints: (57). Finally, the upper-level model formulates capacity demand curves based on the GenCos’ investment obtained from the middle and lower level problems through an iterative algorithm.

\subsection {The solving method for the capacity demand curve formulation model}

With the reformulation of the middle and lower level models, the tri-level capacity demand curve formulation model could be solved with the following process:

\textit{\textbf{Step 1:}}Initialize the generation investment capacity $P_{j,k}^{\Delta }(0)$ to 0, set the upper limits for the capacities that can be provided to the other areas and those of generation capability and flexibility and resilience $P_{i,i^{\prime}}^{LC}(0), P_{i,i^{\prime}}^{LF}(0)$ to zero.

\textit{\textbf{Step 2:} }Solve the upper-level capacity demand curve formulation model (7)-(25), and obtain the capacity demand curves in the m-th iteration. The demand curves comprise a series of demand-price points acquired by solving model (7)-(25) with the capacity demands at inflection points.

\textit{\textbf{Step 3:}} Solve the generation investment equilibrium composed by the middle and lower level as stated in Section IV subsection A, obtain the generation investment capacities $P_{j,k}^{\Delta }(m)$ in the m-th iteration. 

\textit{\textbf{Step 4:}} If $\left\| P_{j,k}^{\Delta }(m)-P_{j,k}^{\Delta }(m-1) \right\|$ is smaller than the convergence criterion, the capacity demand curve formulating results can be determined as the capacity demand curves in m-th iteration. 

\textit{\textbf{Step 5:}} With the decisions of Gencos obtained in \textit{\textbf{Step 3}}, set the capacity that can be provided to the other areas for generation capability and flexibility and resilience $P_{i,i^{\prime}}^{LC}(m), P_{i,i^{\prime}}^{LF}(m)$ in an m-th iteration as $P_{i,i^{\prime}}^{CO}, P_{i,i^{\prime}}^{FO}$. Then go to \textit{\textbf{Step 2}}.

\section {Case study}
\subsection {Data description}

This section conducts case studies by adopting a two-area interconnected power system, the installed capacities shown in Fig.7, to verify the feasibility and effectiveness of the proposed model and methodology. The programming was executed using MATLAB on an i7-8700 desktop with 16G RAM, with GUROBI as the solver.

\begin{figure}
  \begin{center}
  \includegraphics[width=3.5in]{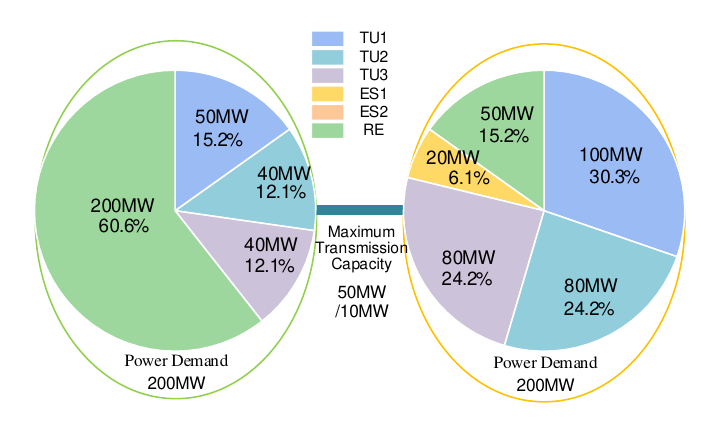}\\
  \caption{The two-area interconnected power system}\label{circuit_diagram}
  \end{center}
\end{figure}

The annual investment cost, maximum generation investment capacity, and several performance parameters of each type of unit are illustrated in Table I. The maximum market clearing prices $\alpha _{\max }^{E}$ in energy market is set as 90 €/MW, and $\alpha _{k,\max }^{CP}, \alpha _{k,\max }^{CF}$ both are set as 260 €/MW.d for all types of units. The charging cost of ES is 20 €/MW. The limit of RoCoF is 2 Hz/s \cite{Uijlings2013}. ${{\pi }_{i}}$ obeys the left-skewed probability distribution of average value as 0.01, standard deviation as 0.015, skewness value as 1.5 and kurtosis value as 3. ${{T}^{B}}$ is 0.5 h.
We generated 12 scenarios using typical wind, solar, and load curves in 12 months and used these 12 scenarios in the upper level model (Obj: (7), s.t. (8)- (25)). Typical curves of the four seasons were selected for model (48) - (54). In this work, three Gencos are set up, among which Genco K1 holds the TU 3 and all ESs in Area 1. The RE, TU 1, TU 2 in Area 1, and all TUs in Area 2 belong to Genco K2. Genco K3 holds the RE and all ESs in Area 2. ${{\partial }_{K1}}, {{\partial }_{K3}}$ is set as 0.7, while ${{\partial }_{K2}}$ is 0.5.

\subsection {Fundamental results}

\begin{table*}[!t]
\caption{Parameters of each type of unit\label{tab:table1}}
\centering
\begin{tabular}{ccccccc}
\hline
Type of unit & RE & TU1 & TU2 & TU3 & ES1 & ES2 \\
\hline
Investment cost(€/kW)/ dura1tion  (year) & 650/25 & 780/30  & 1040/35 & 1300/40  & 1040/25 & 1300/20 \\

Maximum investment capacity area 1/ area 2 (MW) & 60/30 & 50/80 &50/80 & 50/80 & 10/10 & 10/10 \\

Coefficient of confidence capacity  & 0.4 &1  & 1 &1 & 0.6 & 0.5 \\

Ramping performance coefficient  & / & 0.6 & 0.65 & 0.8 &0.9& 1 \\

Inertia constant (s)  & / & 4  & 4 & 8 & 10  & 0 \\

Available time coefficient (h)  & / & 0.13 & 0.16 & 0.33 & 0.45 & 0.48 \\
\hline
\end{tabular}
\end{table*}

This subsection illustrates the fundamental results of formulating the capacity demand curves and how they affect the generation investment. Two cases are adopted, with the inter-area transmission capacity set at 50 MW (Case 1) and 10MW (Case 2) to represent the cases in which the inter-area transmission capacity is abundant (Case 1) and deficient (Case 2), respectively. 

\begin{figure}
  \begin{center}
  \includegraphics[width=2.5in]{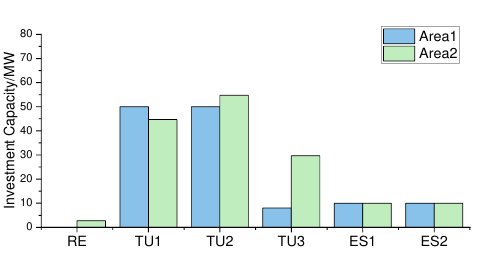}\\
  \caption{The investment capacity of Gencos in Case 1}\label{circuit_diagram}
  \end{center}
\end{figure}

\begin{figure}
  \begin{center}
  \includegraphics[width=2.5in]{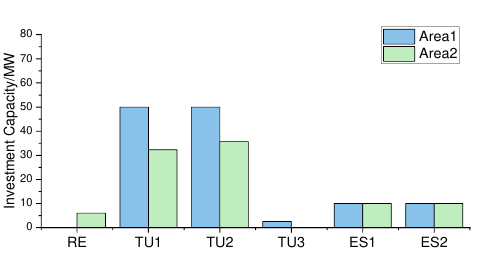}\\
  \caption{The investment capacity of Gencos in Case 2}\label{circuit_diagram}
  \end{center}
\end{figure}

Fig.8 and Fig.9 show the generation investment capacities in two areas in Case 1 and Case 2. Firstly, in Case 1, the investment capacity of TU1 and TU2 in Area 1 reaches the upper limit of 50MW, while that in Area 2 is 44.7MW and 54.7MW. This is attributed to the fact that the TU has a significant advantage in the energy market due to its lower marginal cost \cite{Yang2019}, and the existing installed capacity of TU in Area 1 is lower than that in Area 2, so there is higher profitability for investment in TU in Area 1. In addition, due to the high capacities of TUs in Area 2 and the abundant transmission capacity between the two areas, sufficient capacity for flexibility and resilience can be acquired in Case 1. Therefore, the invested capacities of TUs are lower in Area 1, with only 8MW of TU 3. On the contrary, in Case 2, the capacity for flexibility and resilience that can be provided from Area 2 to Area 1 is limited due to the deficient transmission capacity between the two areas, leading to the investment on TUs in Area 1 significantly larger compared to Area 2, specifically the investment capacity on TU 1 and TU2 reaches the upper limit of 50MW, with an additional investment capacity of 2.6MW on TU 3, while the invested capacity on TUs in Area 2 is 32.3MW, 35.7MW and 0MW respectively. At the same time, the investment capacity of RE in Area 2 increases to 6.0MW compared to Case 1. This demonstrates that the proposed capacity market can effectively guide Gencos in investing in appropriate types of units to satisfy the demand for flexibility and resilience.

\begin{table*}[!t]
\caption{Provisioned capacity for generation capability and flexibility and resilience in Case 1\label{tab:table1}}
\centering
\begin{tabular}{cccccccc}
\hline
& Type of unit& RE & TU1 & TU2 & TU3 & ES1 & ES2 \\
\hline
area 1& Provisioned capacity for generation capability (MW)& 24.28 & 23.04  & 23.03 &23.05& 6.00  & 0.00 \\

&Provisioned capacity for flexibility and resilience (MW)  &0.00 & 9.09 &16.26 &6.65 & 0.00 &5.00 \\

area 2& Provisioned capacity for generation capability (MW) & 21.11 &25.26  & 25.27 &25.26 & 11.88 & 0.00 \\

& Provisioned capacity for flexibility and resilience (MW)&0.00 &5.43 & 9.30& 4.97 & 6.12 & 5.00 \\
\hline
\end{tabular}
\end{table*}

\begin{table*}[!t]
\caption{Provisioned capacity for generation capability and flexibility and resilience in Case 2\label{tab:table1}}
\centering
\begin{tabular}{cccccccc}
\hline
 &Type of unit & RE & TU1 & TU2 & TU3 & ES1 & ES2 \\
\hline
area 1& Provisioned capacity for generation capability (MW)& 25.45 & 24.70  & 24.70  &24.70  &6.00   & 0.00 \\

& Provisioned capacity for flexibility and resilience (MW) & 0.00  &13.44  &19.67  &10.34 & 0.00  &5.00 \\

area 2& Provisioned capacity for generation capability (MW) & 20.00  &91.67   & 0.00  &0.00  & 0.00   & 0.00 \\

 &Provisioned capacity for flexibility and resilience (MW) & 0.00  & 10.26 & 9.40  &13.67 & 12.00 & 5.00 \\
\hline
\end{tabular}
\end{table*}

Tables II and III list the provisioned capacities for generation capability, flexibility and resilience in Case 1 and Case 2. The overall provisioned capacities for generation capability, ramping capability, inertia and recovery capability can be seen in Fig.10 and Fig.11. In Case 1, it shows that the sum of provisioned capacities and transmission capacity for ramping capability and inertia and recovery capability of Area 2 are larger than those in Area 1, due to the larger demand curve price on flexibility and resilience. Specifically, Area 1 purchases ramping capability of 11.77MW, inertia of 49.5MW.s and energy supply for recovery capability of 3.7MWh. In contrast, in Area 2, there is only ramping capability of 38.35 MW, inertia of 269.5MW.s, and energy supply for recovery capability of 13.96 MWh, as shown in Fig.10.

\begin{figure*}
  \begin{center}
  \includegraphics[width=5in]{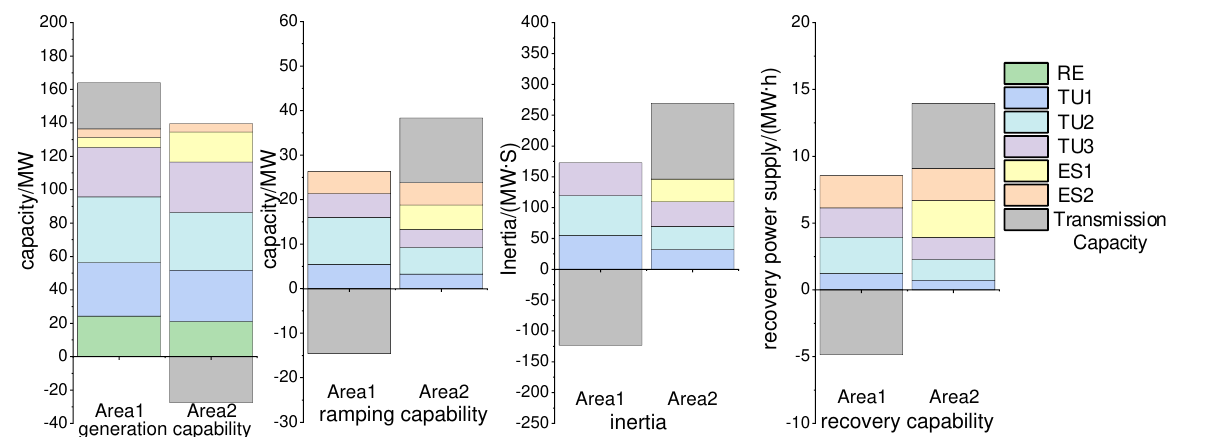}\\
  \caption{The overall provisioned generation capability, ramping capability, inertia and recovery capability in Case 1}\label{circuit_diagram}
  \end{center}
\end{figure*}

\begin{figure*}
  \begin{center}
  \includegraphics[width=5in]{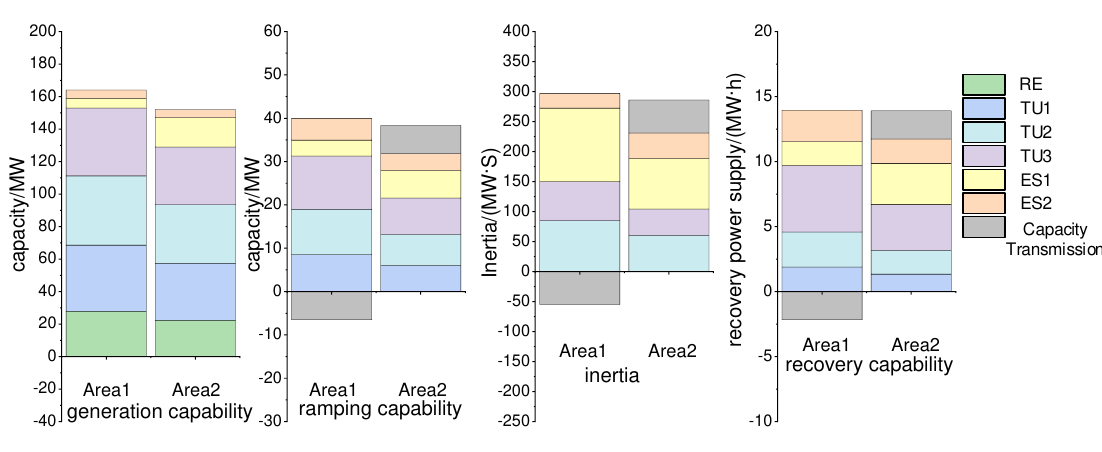}\\
  \caption{The overall provisioned generation capability, ramping capability, inertia and recovery capability in Case 2}\label{circuit_diagram}
  \end{center}
\end{figure*}

To further study the guidance effect of the proposed capacity market on the generation investment of GenCos, the capacity demand curves of two areas in Case 2 are discussed, as shown in Fig.12. Note that for an intuitive and numerical representation of the demand for flexibility and resilience,  the capacity demand for ramping capability, inertia, and recovery capability is transformed into corresponding capabilities with the average ramping performance, the inertia constant, and the available time coefficient.

\begin{figure*}
  \begin{center}
  \includegraphics[width=7in]{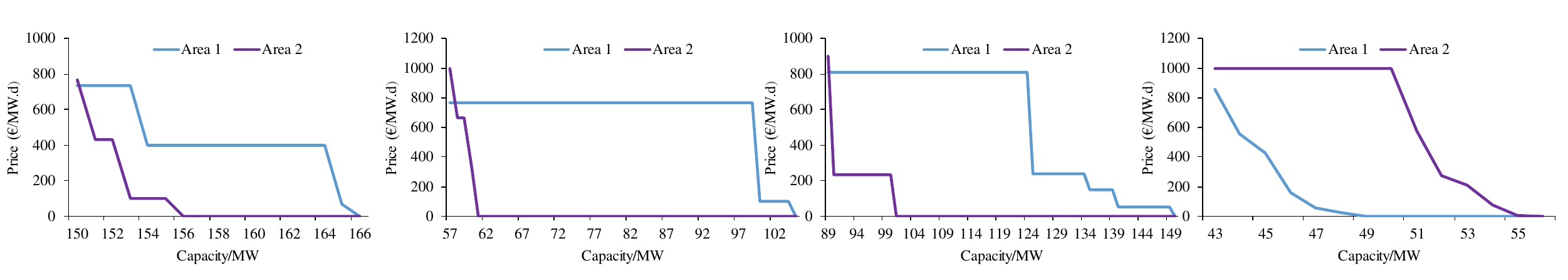}\\
  \caption{Capacity demand curves in Case 2
(a) Generation capability  (b) Ramping capability (c) Inertia (d) Recovery capability}\label{circuit_diagram}
  \end{center}
\end{figure*}

\begin{figure*}
  \begin{center}
  \includegraphics[width=7in]{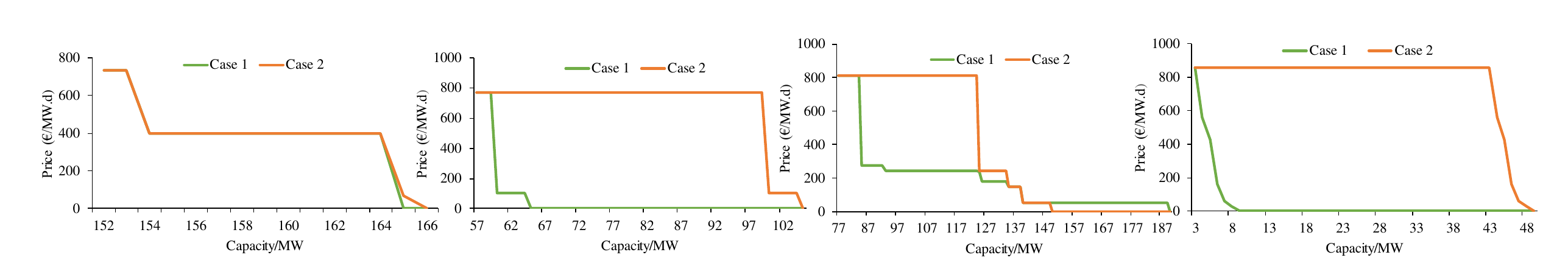}\\
  \caption{Capacity demand curves in Area 1 
(a) Generation capability (b) Ramping capability (c) Inertia (d) Recovery capability
}\label{circuit_diagram}
  \end{center}
\end{figure*}

\begin{figure*}
  \begin{center}
  \includegraphics[width=7in]{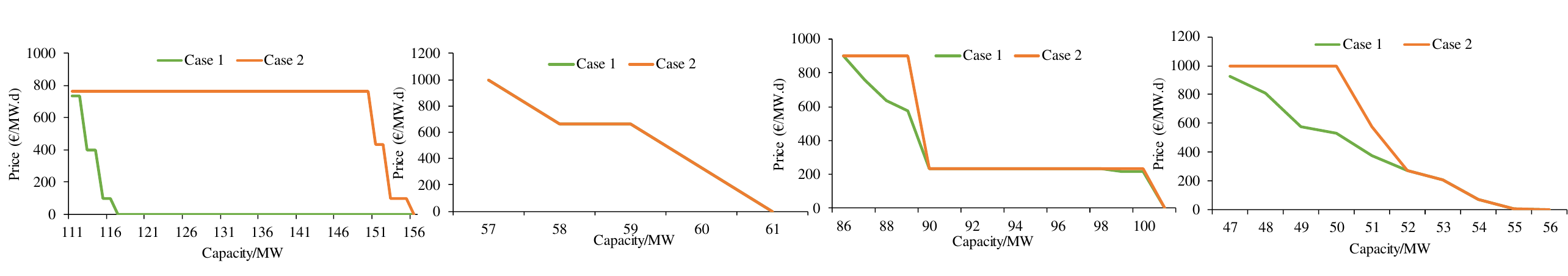}\\
  \caption{Capacity demand curves in Area 2
(a) Generation capability  (b) Ramping capability (c) Inertia (d) Recovery capability}\label{circuit_diagram}
  \end{center}
\end{figure*}

Firstly, the maximum demand on the capacity demand curve for generation capability in the two areas is similar (156MW in Area 1 and 166MW in Area 2), as the total installed capacity and load demand are close in these two areas. As for the ramping capability and inertia, which are closely related to the installed capacity of RE, the maximum demand on the capacity demand curve for these two capabilities in Area 1 (105 MW and 150 MW in Case 2) is higher than that in Area 2 (62 MW and 101MW in Case 2), as the RE is mainly installed in Area 1. The differences in the capacity demand curves of ramping capability and inertia indicate that the proposed capacity market can effectively reflect the demand for flexibility in each area. Regarding the recovery capability, the capacity demand curves of the two areas are also similar, as the demand for recovery capability is relatively small. 

In addition, as shown in Fig.13, the change of inter-area transmission capacity significantly impacts the capacity demand curves of ramping, inertia and recovery capability in Area 1. Due to the high installed capacity of RE and low installed capacity of TUs and ESs in Area 1, when the transmission capacity between the two areas is limited, the demand for flexibility and resilience in Area 1 can hardly be supported by Area 2. Therefore, the provisioned capacities for ramping capability, inertia and recovery capability need to increase. Thus, the prices in the demand curve for flexibility and resilience capability increase in Case 2 compared to those in Case 1. These changes in capacity demand and capacity price on the inertia capacity demand curve effectively promote the investment in TUs in Area 1.

As shown in Fig.14, the change of inter-area transmission capacity also significantly impacts the capacity demand curve of generation capability in Area 2. The deficiency of generation capability in Area 2 is due to the limited capacity of existing generation units, and when the inter-area transmission capacity is deficient, insufficient generation capability can be provided within Area 2, and the prices on the demand curve could be higher. In addition, as shown in Fig.10 and Fig.11, different from Area 1, the demand for generation capability has been increased from 112MW to 152MW in Area 2. This also indicates the insufficiency of generation capability in Area 2.

In summary, each area could formulate the capacity demand curves for generation capability, ramping capability, inertia, and recovery capability to reflect the surplus or insufficiency of these capabilities by adjusting the slope of the capacity demand curve and the corresponding intercept. Such capacity demand curves could help meet the demand for generation capability, flexibility, and resilience and guide the proper investment of GenCos.

\subsection {The comparison with the case without incorporating the flexibility and resilience demand}

This subsection presents a comparative analysis of Case 3, in which flexibility and resilience demands are not incorporated. The inter-area transmission capacity in Case 3 is still set to 50MW. The generation investment capacity under the market mechanism, without incorporating the flexibility and resilience demands, is shown in Fig.15. 

By comparing Fig.15 with Fig.8, it is evident that incorporating flexibility and resilience demands can effectively stimulate the investment in TUs in Area 2. In Case 1, where the capacity market includes flexibility and resilience demands, the investment capacity of TUs in Area 2 reaches 129MW. In contrast, in Case 3, where these demands are excluded, the investment capacity of TUs in Area 2 is only 68MW.

\begin{figure}
  \begin{center}
  \includegraphics[width=2.5in]{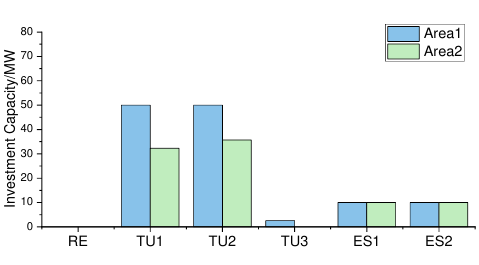}\\
  \caption{The generation investment capacity of GenCos without incorporating the flexibility and resilience demand into the capacity market}\label{circuit_diagram}
  \end{center}
\end{figure}

\begin{figure}
  \begin{center}
  \includegraphics[width=2.5in]{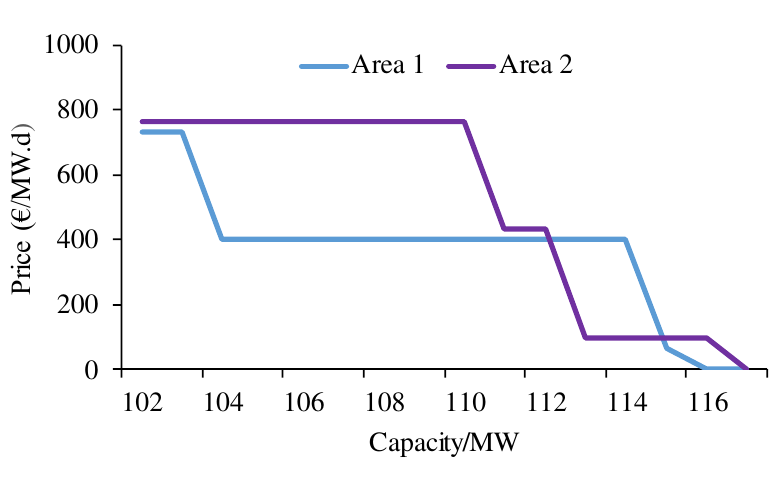}\\
  \caption{Capacity demand curves without incorporating the flexibility and resilience demand}\label{circuit_diagram}
  \end{center}
\end{figure}

\begin{table*}[!t]
\caption{Provisioned capacity for generation capability without incorporating the flexibility and resilience demand\label{tab:table1}}
\centering
\begin{tabular}{cccccccc}
\hline
&Type of unit & RE & TU1 & TU2 & TU3 & ES1 & ES2 \\
\hline
Area 1 &Provisioned capacity for generation capability (MW)
 & 14.54 & 13.03 & 13.03 &13.04 &3.00 & 0.00 \\

& Provisioned capacity for flexibility and resilience (MW)
&0.00 &13.03&15.05   &13.03 & 3.00 &5.00\\

Area 2 &Provisioned capacity for generation capability (MW)
& 14.53 &13.04 &13.03 &13.04  & 8.24   & 2.50 \\

& Provisioned capacity for flexibility and resilience (MW) 
& 0.00 & 14.53 &16.04 &13.03 & 9.76 &2.50 \\
\hline
\end{tabular}
\end{table*}

\begin{figure*}
  \begin{center}
  \includegraphics[width=5in]{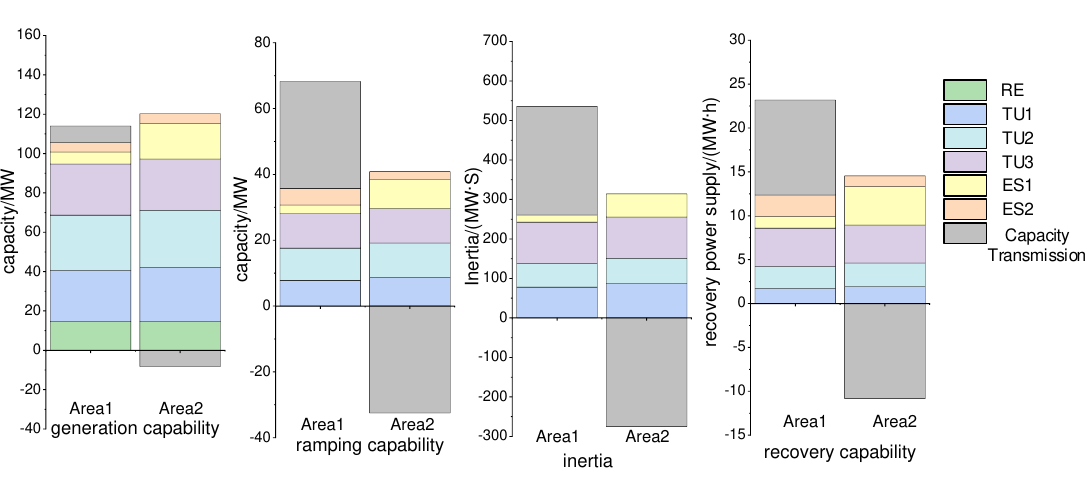}\\
  \caption{The overall provisioned generation capability and the availability of ramping capability, inertia and recovery capability without incorporating the flexibility and resilience demand}\label{circuit_diagram}
  \end{center}
\end{figure*}

Fig.16 displays the capacity demand curves for generation capability in Areas 1 and 2 of Case 3. Table IV lists the provisioned capacity for generation capability, flexibility, and resilience. Fig. 17 shows the availability of ramping, inertia, and recovery capability when the flexibility and resilience demands are not incorporated (Case 3). In Fig.16 and Fig.17, the provisioned generation capacities in Area 1 are similar to those in Area 2, as the prices in the generation capacity market are comparable for both areas. Area 1 won 114 MW in the generation capacity market, while Area 2 won 112 MW. However, when comparing with Fig.10, which includes the flexibility and resilience products (Case 1), it is clear that more capabilities for flexibility and resilience would be provisioned in Area 2 (38.35MW  of ramping, 269.5MWs of inertia and 13.96MWh of recovery capability, as shown in Fig.10). In contrast, in Case 3, there are only 8 MW of ramping, 39MWs of inertia and 4MWh of recovery capability, as shown in Fig.17. 

\begin{table}[!t]
\caption{Comparisons on market results\label{tab:table1}}
\centering
\begin{tabular}{ccccc}
\hline
 & capacity  & economic  & economic& Overall \\
Cases&  market  & losses &  losses  & cost (€)\\
 & cost (€) &  in area 1 (€) &  in area 2 (€) &\\
\hline
case1($*10^7$) & 18.01 & 7.88 & 6.68 &32.57\\

case3($*10^7$) & 15.20 & 1.16 & 19.16 &35.53\\
\hline
\end{tabular}
\end{table}

Table V presents the market results for the two different formulations of the capacity demand curve and the economic losses resulting from insufficiently provisioned capacities in each area. The purchasing cost of the capacity market is lower when flexibility and resilience demands are not incorporated (€$15.20*10^7$ in Case 3). However, the expected economic losses due to insufficient capacity provision in Area 2 (€$19.16*10^7$) are higher compared to when flexibility and resilience demands are incorporated (€$6.68*10^7$). This is because TUs, which provide ramping, inertia, and recovery capability, invest less in Area 2 when flexibility and resilience demands are not considered. The expected economic losses in Area 1 are lower in Case 3 (€$1.16*10^7$), thanks to the provisioned capacities transmitted from Area 2. Overall, incorporating flexibility and resilience demands in the capacity market can significantly reduce the economic losses associated with insufficient flexibility and resilience capabilities.

\subsection{The necessity of formulating capacity demand curves by each area}
This subsection compares the results of formulating the capacity demand curves for the two areas separately (Case 1) and as a whole (Case 4). Fig.18 illustrates the generation investment capacities when the capacity demand curves are developed for both areas collectively. The provisioned capacities for generation capability, flexibility and resilience are listed in Table VI. The demand curves and the corresponding provisioned capacities for generation capability, ramping capability, inertia, and recovery capability are shown in Fig.19 and Fig.20.

\begin{figure}
  \begin{center}
  \includegraphics[width=3in]{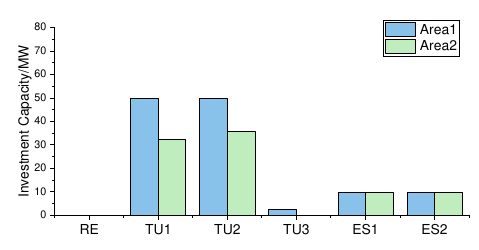}\\
  \caption{The investment capacity of GenCos when the demand curves are formulated by taking the two areas as a whole}\label{circuit_diagram}
  \end{center}
\end{figure}

\begin{figure*}
  \begin{center}
  \includegraphics[width=7.2in]{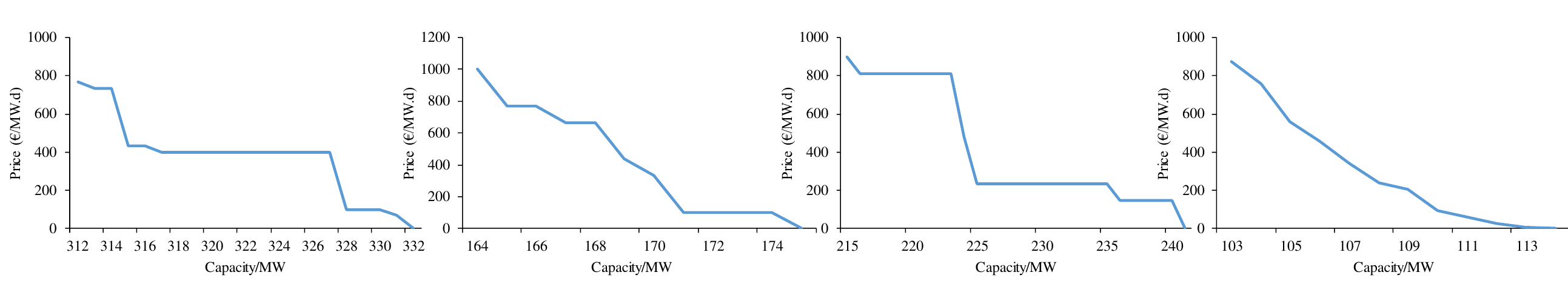}\\
  \caption{The capacity demand curves when the demand curves are formulated by taking the two areas as a whole (a) Generation capability  (b) Ramping capability (c) Inertia (d) Recovery capability}\label{circuit_diagram}
  \end{center}
\end{figure*}

\begin{table*}[!t]
\caption{Provisioned capacity for generation capability and for flexibility and resilience when the demand curves are formulated by taking the two areas as a whole\label{tab:table1}}
\centering
\begin{tabular}{cccccccc}
\hline
&Type of unit & RE & TU1 & TU2 & TU3 & ES1 & ES2 \\
\hline
Area 1 &Provisioned capacity for generation capability (MW)
 & 23.37 & 21.68 & 21.67 &21.67 &3.00 & 0.00 \\

& Provisioned capacity for flexibility and resilience (MW)
&0.00 &0.66&58.29   &0.00 & 6.00 &5.00\\

Area 2 &Provisioned capacity for generation capability (MW)
& 20.00 &21.66 &21.69 &21.69  & 0.00   & 0.00 \\

& Provisioned capacity for flexibility and resilience (MW) 
& 0.00 & 1.71 &58.91 &0.00 & 18.00 &5.00 \\
\hline
\end{tabular}
\end{table*}

\begin{figure*}
  \begin{center}
  \includegraphics[width=5in]{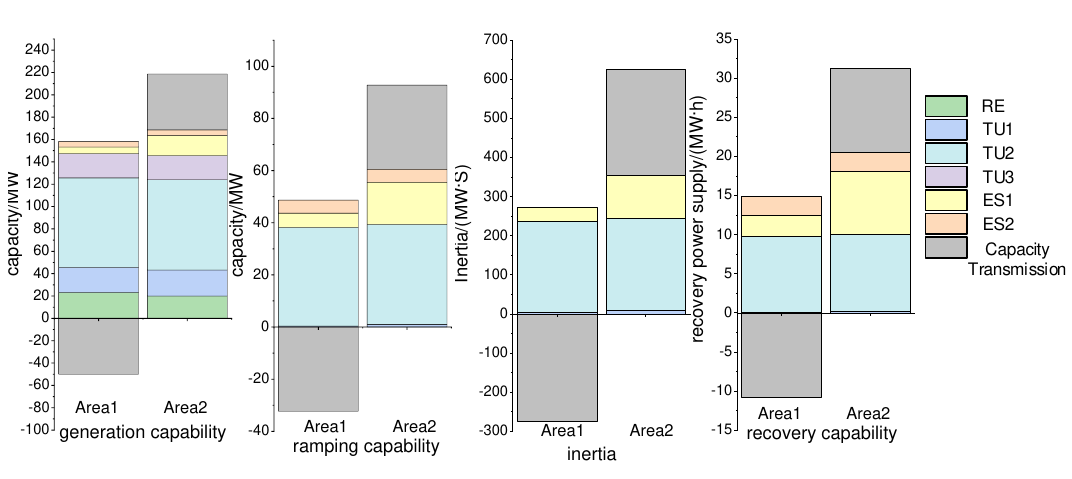}\\
  \caption{The overall provisioned generation capability, ramping capability, inertia and recovery capability when the demand curves are formulated by taking the two areas as a whole}\label{circuit_diagram}
  \end{center}
\end{figure*}

\begin{table}[!t]
\caption{Comparisons on market results\label{tab:table1}}
\centering
\begin{tabular}{c c c c c}
\hline
 & capacity & economic   & economic & Overall\\
Cases& market &losses in&losses in  &cost (€)\\
 & cost (€) &   area 1 (€) &  area 2 (€) & \\
\hline
case1-50MW($*10^7$)& 18.01 & 7.88 & 6.68 & 32.57\\

case4-50MW($*10^7$)& 17.90 & 10.70 & 5.82 & 34.42\\

case2-10MW($*10^7$)& 20.70 & 0.73 & 7.54 & 28.97\\

case4-10MW($*10^7$)& 17.90 & 0.89 & 10.20 & 28.99\\
\hline
\end{tabular}
\end{table}

When the capacity demand curves are formulated for the two areas as a whole, the total investment capacities of TUs and ESs are 123MW and 88MW in Areas 1 and 2, respectively. The capacity could satisfy the demand for flexibility and resilience in both areas. However, the risk of capacity inadequacy might arise if the transmission capacity is insufficient. Table VII provides the market results and the expected losses due to inadequate provisioned capacity with varying transmission capacities between the two areas.

The large amount of TUs in Area 1 can provide ramping capability, inertia, and recovery capability for the whole system, resulting in a lower overall capacity demand when the demand curves are formulated for both areas together. Consequently, the purchasing cost of the capacity market (€$17.90*10^7$ in Case 4) is reduced compared to the case when the demand curves are developed separately for each area (€$18.01*10^7$ in Case 1). However, the expected economic losses due to insufficient provisioned capacity in Area 1 increase when formulating the demand curves for both areas collectively (€$10.70*10^7$).

When the transmission capacity between two areas is inadequate (Case 2 and Case 4 with 10MW), the expected economic losses due to insufficient provisioned capacity in Area 2 (€$10.20*10^7$) rise more significantly, as the investment and the provisioned capacities are concentrated in Area 1. 

Although the purchasing cost of capacity market increases when the capacity demand curves are formulated separately for each area, the impact on guiding generation investment by GenCos and reducing the expected economic losses caused by insufficient provisioned capacity is improved. Therefore, we recommend formulating the capacity demand curves for each area of the interconnected power system separately when designing the capacity market, ensuring that the demands for generation capability, flexibility and resilience are all met.

\section{Conclusion}
Incorporating flexibility and resilience demand into the capacity market could stimulate investment and guarantee the adequacy of the resources that provide flexibility and resilience to the power systems facing the rapidly increasing proportion of RE. This paper first specifies the flexibility and resilience demand into three trading products: ramping, inertia, and recovery capability, and proposes a modified capacity market framework. Then, a tri-level model is established to formulate the capacity demand curves, with the middle and lower level problems seeking the generation investment equilibrium among GenCos participating in both capacity and energy markets. A Nash-Cournot model with an equivalent quadratic programming formulation is proposed to solve the generation investment equilibrium, and an iterative algorithm is devised to coordinate the capacity demand curve formulation problem and the generation investment equilibrium problem.

The case studies indicate that: 1) to incorporate the flexibility and resilience demand into the capacity market can effectively reflect the scarcity or surplus of the generation capability, ramping capability, inertia and recovery capability, as well as guiding the generation investment of GenCos to ensure resource adequacy. 2) Incorporating flexibility and resilience demand into the capacity market leads to higher purchasing costs in the capacity market, while a significant reduction in the expected economic losses caused by insufficient provisioned capabilities for flexibility and resilience. 3) The capacity demand curves formulated for each area individually in an interconnected power system can lead to higher purchasing costs in the capacity market. However, multi-region interconnections are generally connected by only a limited number of high-voltage lines. In the case of a major failure, the transmission capacity of these interconnection lines could be substantially reduced. In contrast to formulating capacity demand curves for the entire interconnected system, developing them for each area separately offers notable advantages in ensuring the necessary flexibility and resilience for the adequacy of each region.

\ifCLASSOPTIONcaptionsoff
  \newpage
\fi




\bibliographystyle{IEEEtran}
\bibliography{IEEEabrv,Bibliography}

\begin{thebibliography}{10}
\providecommand{\url}[1]{#1}
\csname url@samestyle\endcsname
\providecommand{\newblock}{\relax}
\providecommand{\bibinfo}[2]{#2}
\providecommand{\BIBentrySTDinterwordspacing}{\spaceskip=0pt\relax}
\providecommand{\BIBentryALTinterwordstretchfactor}{4}
\providecommand{\BIBentryALTinterwordspacing}{\spaceskip=\fontdimen2\font plus
\BIBentryALTinterwordstretchfactor\fontdimen3\font minus \fontdimen4\font\relax}
\providecommand{\BIBforeignlanguage}[2]{{%
\expandafter\ifx\csname l@#1\endcsname\relax
\typeout{** WARNING: IEEEtran.bst: No hyphenation pattern has been}%
\typeout{** loaded for the language `#1'. Using the pattern for}%
\typeout{** the default language instead.}%
\else
\language=\csname l@#1\endcsname
\fi
#2}}
\providecommand{\BIBdecl}{\relax}
\BIBdecl

\bibitem{Guerra2022}
K.~Guerra, P.~Haro, R.~E. Guti{\'{e}}rrez, and A.~G{\'{o}}mez-Barea, ``{Facing the high share of variable renewable energy in the power system: Flexibility and stability requirements},'' \emph{Appl. Energy}, vol. 310, 2022.

\bibitem{A.E.M.Operator2016}
\BIBentryALTinterwordspacing
{Ausralian Energy Market Operator (AEMO)}, ``{Black System South Australia 28 September 2016},'' Tech. Rep. SEPTEMBER 2016, 2017. [Online]. Available: \url{https://www.aemo.com.au/-/media/Files/Electricity/NEM/Market_Notices_and_Events/Power_System_Incident_Reports/2017/Integrated-Final-Report-SA-Black-System-28-September-2016.pdf}
\BIBentrySTDinterwordspacing

\bibitem{Shiltz2016}
D.~J. Shiltz, M.~Cvetkovi{\'{c}}, and A.~M. Annaswamy, ``{An Integrated Dynamic Market Mechanism for Real-Time Markets and Frequency Regulation},'' \emph{IEEE Trans. Sustain. Energy}, vol.~7, no.~2, pp. 875--885, 2016.

\bibitem{Silva-Rodriguez2024}
L.~Silva-Rodriguez, A.~Sanjab, E.~Fumagalli, and M.~Gibescu, ``{Light robust co-optimization of energy and reserves in the day-ahead electricity market},'' \emph{Appl. Energy}, vol. 353, no.~A, 2024.

\bibitem{Cui2018}
M.~Cui and J.~Zhang, ``{Estimating ramping requirements with solar-friendly flexible ramping product in multi-timescale power system operations},'' \emph{Appl. Energy}, vol. 225, pp. 27--41, 2018.

\bibitem{Qiu2024}
D.~Qiu, A.~M. Baig, Y.~Wang, L.~Wang, C.~Jiang, and G.~Strbac, ``{Market design for ancillary service provisions of inertia and frequency response via virtual power plants: A non-convex bi-level optimisation approach},'' \emph{Appl. Energy}, vol. 361, 2024.

\bibitem{Panteli2017}
M.~Panteli, D.~N. Trakas, P.~Mancarella, and N.~D. Hatziargyriou, ``{Power Systems Resilience Assessment: Hardening and Smart Operational Enhancement Strategies},'' \emph{Proc. IEEE}, vol. 105, no.~7, pp. 1202--1213, 2017.

\bibitem{Ma2013}
J.~Ma, V.~Silva, R.~Belhomme, D.~S. Kirschen, and L.~F. Ochoa, ``{Evaluating and planning flexibility in sustainable power systems},'' \emph{IEEE Trans. Sustain. Energy}, vol.~4, no.~1, pp. 200--209, 2013.

\bibitem{Zhu2024}
Y.~Zhu, Y.~Xiao, X.~Wang, C.~Chen, Z.~Lu, and X.~Wang, ``{Enhancing Distribution System Resilience With Peer-to-Peer Transactions},'' \emph{IEEE Trans. Power Syst.}, vol.~40, no.~1, pp. 907--919, 2025.

\bibitem{Bowring2013}
J.~Bowring, ``{Capacity marketsin PJM},'' \emph{Econ. Energy Environ. Policy}, vol.~2, no.~2, pp. 47--64, 2013.

\bibitem{Harbord2014}
D.~Harbord and M.~Pagnozzi, ``{Britain's electricity capacity auctions: Lessons from Colombia and New England},'' \emph{Electr. J.}, vol.~27, no.~5, pp. 54--62, 2014.

\bibitem{Khan2018}
A.~S.~M. Khan, R.~A. Verzijlbergh, O.~C. Sakinci, and L.~J. {De Vries}, ``{How do demand response and electrical energy storage affect (the need for) a capacity market?}'' \emph{Appl. Energy}, vol. 214, pp. 39--62, 2018.

\bibitem{Wang2024}
R.~Wang, S.~Wang, G.~Geng, and Q.~Jiang, ``{Multi-time-scale capacity credit assessment of renewable and energy storage considering complex operational time series},'' \emph{Appl. Energy}, vol. 355, 2024.

\bibitem{Hobbs2007}
B.~F. Hobbs, M.~C. Hu, J.~G. I{\~{n}}{\'{o}}n, S.~E. Stoft, and M.~P. Bhavaraju, ``{A dynamic analysis of a demand curve-based capacity market proposal: The PJM reliability pricing model},'' \emph{IEEE Trans. Power Syst.}, vol.~22, no.~1, pp. 3--14, 2007.

\bibitem{Byers2018}
C.~Byers, T.~Levin, and A.~Botterud, ``{Capacity market design and renewable energy: Performance incentives, qualifying capacity, and demand curves},'' \emph{Electr. J}, vol.~31, no.~1, pp. 65--74, 2018.

\bibitem{Zhao2017}
F.~Zhao, T.~Zheng, and E.~Litvinov, ``{Constructing Demand Curves in Forward Capacity Market},'' \emph{IEEE Trans. Power Syst.}, vol.~33, no.~1, pp. 525--535, 2017.

\bibitem{Kaminski2023}
S.~Kaminski, K.~Bruninx, and E.~Delarue, ``{A Comparative Study of Capacity Market Demand Curve Designs Considering Risk-Averse Market Participants},'' \emph{IEEE Trans. Energy Mark. Policy Regul.}, vol.~1, no.~4, pp. 286--296, 2023.

\bibitem{Fang2021}
X.~Fang, Q.~Hu, R.~Bo, and F.~Li, ``{Redesigning capacity market to include flexibility via ramp constraints in high-renewable penetrated system},'' \emph{Int. J. Electr. Power Energy Syst.}, vol. 128, 2021.

\bibitem{Hu2023}
J.~Hu, Z.~Yan, X.~Xu, and S.~Chen, ``{Inertia Market: Mechanism Design and Its Impact on Generation Mix},'' \emph{J. Mod. Power Syst. Clean Energy}, vol.~11, no.~3, pp. 744--756, 2023.

\bibitem{Liang2023}
Z.~Liang, R.~Mieth, and Y.~Dvorkin, ``{Inertia Pricing in Stochastic Electricity Markets},'' \emph{IEEE Trans. Power Syst.}, vol.~38, no.~3, pp. 2071--2084, 2023.

\bibitem{Saraf2009}
N.~Saraf, K.~McIntyre, J.~Dumas, and S.~Santoso, ``{The annual black start service selection analysis of ERCOT grid},'' \emph{IEEE Trans. Power Syst.}, vol.~24, no.~4, pp. 1867--1874, 2009.

\bibitem{Grimm2021}
V.~Grimm, B.~R{\"{u}}ckel, C.~S{\"{o}}lch, and G.~Z{\"{o}}ttl, ``{The impact of market design on transmission and generation investment in electricity markets},'' \emph{Energy Econ.}, vol.~93, 2021.

\bibitem{Bhagwat2017}
P.~C. Bhagwat, A.~Marcheselli, J.~C. Richstein, E.~J. Chappin, and L.~J. {De Vries}, ``{An analysis of a forward capacity market with long-term contracts},'' \emph{Energy Policy}, vol. 111, pp. 255--267, 2017.

\bibitem{Hach2016}
D.~Hach, C.~K. Chyong, and S.~Spinler, ``{Capacity market design options: A dynamic capacity investment model and a GB case study},'' \emph{Eur. J. Oper. Res.}, vol. 249, no.~2, pp. 691--705, 2016.

\bibitem{Wogrin2013}
S.~Wogrin, J.~Barqu{\'{i}}n, and E.~Centeno, ``{Capacity expansion equilibria in liberalized electricity markets: An EPEC approach},'' \emph{IEEE Trans. Power Syst.}, vol.~28, no.~2, pp. 1531--1539, 2013.

\bibitem{Mousavian2020}
S.~Mousavian, A.~J. Conejo, and R.~Sioshansi, ``{Equilibria in investment and spot electricity markets: A conjectural-variations approach},'' \emph{Eur. J. Oper. Res.}, vol. 281, no.~1, pp. 129--140, 2020.

\bibitem{Hoschle2018}
H.~Hoschle, H.~{Le Cadre}, Y.~Smeers, A.~Papavasiliou, and R.~Belmans, ``{An ADMM-Based Method for Computing Risk-Averse Equilibrium in Capacity Markets},'' \emph{IEEE Trans. Power Syst.}, vol.~33, no.~5, pp. 4819--4830, 2018.

\bibitem{Ihlemann2022}
M.~Ihlemann, A.~van Stiphout, K.~Poncelet, and E.~Delarue, ``{Benefits of regional coordination of balancing capacity markets in future European electricity markets},'' \emph{Appl. Energy}, vol. 314, 2022.

\bibitem{Li2022}
R.~Li, B.~J. Tang, B.~Yu, H.~Liao, C.~Zhang, and Y.~M. Wei, ``{Cost-optimal operation strategy for integrating large scale of renewable energy in China's power system: From a multi-regional perspective},'' \emph{Appl. Energy}, vol. 325, 2022.

\bibitem{Cepeda2018}
M.~Cepeda, ``{Assessing cross-border integration of capacity mechanisms in coupled electricity markets},'' \emph{Energy Policy}, vol. 119, pp. 28--40, 2018.

\bibitem{Meyer2015}
R.~Meyer and O.~Gore, ``{Cross-border effects of capacity mechanisms: Do uncoordinated market design changes contradict the goals of the European market integration?}'' \emph{Energy Econ.}, vol.~51, pp. 9--20, 2015.

\bibitem{Brunner2020}
C.~Brunner, G.~Deac, S.~Braun, and C.~Z{\"{o}}phel, ``{The future need for flexibility and the impact of fluctuating renewable power generation},'' \emph{Renew. Energy}, vol. 149, pp. 1314--1324, 2020.

\bibitem{Bie2017}
Z.~Bie, Y.~Lin, G.~Li, and F.~Li, ``{Battling the Extreme: A Study on the Power System Resilience},'' \emph{Proc. IEEE}, vol. 105, no.~7, pp. 1253--1266, 2017.

\bibitem{Makolo2021}
P.~Makolo, R.~Zamora, and T.~T. Lie, ``{The role of inertia for grid flexibility under high penetration of variable renewables - A review of challenges and solutions},'' \emph{Renew. Sustain. Energy Rev.}, vol. 147, 2021.

\bibitem{Wang2017}
Q.~Wang and B.~M. Hodge, ``{Enhancing power system operational flexibility with flexible ramping products: A review},'' \emph{IEEE Trans. Ind. Informatics}, vol.~13, no.~4, pp. 1652--1664, 2017.

\bibitem{Yang2022}
W.~Yang, S.~N. Sparrow, M.~Ashtine, D.~C. Wallom, and T.~Morstyn, ``{Resilient by design: Preventing wildfires and blackouts with microgrids},'' \emph{Appl. Energy}, vol. 313, 2022.

\bibitem{Chicco2004}
G.~Chicco and G.~Gross, ``{Competitive Acquisition of Prioritizable Capacity-Based Ancillary Services},'' \emph{IEEE Trans. Power Syst.}, vol.~19, no.~1, pp. 569--576, 2004.

\bibitem{PJM2022}
\BIBentryALTinterwordspacing
C.~Market and D.~R. Operations, ``{PJM Manual 18 :PJM Capacity Market},'' Tech. Rep., 2021. [Online]. Available: \url{https://wired.pjm.com/-/media/training/nerc-certifications/gen-exam-materials-nov-22-2022/manuals/pjm-capacity-mkt.ashx}
\BIBentrySTDinterwordspacing

\bibitem{S.Oliveira2023}
F.~{S. Oliveira}, B.~William-Rioux, and A.~Pierru, ``{Capacity expansion in liberalized electricity markets with locational pricing and renewable energy investments},'' \emph{Energy Econ.}, vol. 127, 2023.

\bibitem{Hashimoto1985}
\BIBentryALTinterwordspacing
H.~Hashimoto, ``{A Spatial Nash Equilibrium Model},'' in \emph{Spatial Price Equilibrium: Advances in Theory, Computatoon and Applications.}, 1985, pp. 20--40. [Online]. Available: \url{http://link.springer.com/10.1007/978-3-642-46548-2_2}
\BIBentrySTDinterwordspacing

\bibitem{Sohl1985}
\BIBentryALTinterwordspacing
J.~E. Sohl, ``{An Application of Quadratic Programming to the Deregulation of Natural Gas},'' 1985, pp. 196--207. [Online]. Available: \url{http://link.springer.com/10.1007/978-3-642-46548-2_8}
\BIBentrySTDinterwordspacing

\bibitem{Hogan1997}
W.~W. Hogan, ``{A market power model with strategic interaction in electricity networks},'' \emph{Energy J}, vol.~18, no.~4, pp. 107--141, 1997.

\bibitem{Chattopadhyay2004}
D.~Chattopadhyay, ``{Multicommodity Spatial Cournot Model for Generator Bidding Analysis},'' \emph{IEEE Trans. Power Syst}, vol.~19, no.~1, pp. 267--275, 2004.

\bibitem{Uijlings2013}
\BIBentryALTinterwordspacing
{DNV KEMA Energy \& Sustainability} and EirGrid, ``{RoCoF An independent analysis on the ability of Generators to ride through Rate of Change of Frequency values up to 2 Hz/s.}'' EirGrid, London, UK, Tech. Rep. 4478894, 2013. [Online]. Available: \url{http://www.eirgridgroup.com/site-files/library/EirGrid/DNV-KEMA_Report_RoCoF_20130208final_.pdf}
\BIBentrySTDinterwordspacing

\bibitem{Yang2019}
J.~Yang, J.~Zhao, J.~Qiu, and F.~Wen, ``{A Distribution Market Clearing Mechanism for Renewable Generation Units With Zero Marginal Costs},'' \emph{IEEE Trans. Ind. Informatics}, vol.~15, no.~8, pp. 4775--4787, 2019.

\end{thebibliography}
%

\vfill


\end{document}